\documentclass[journal,11pt,draftclsnofoot,onecolumn]{IEEEtran} 

\usepackage[caption=false,font=normalsize,labelfont=rm,textfont=rm]{subfig}
\usepackage{algorithm}
\usepackage{algorithmic}
\usepackage{bm}
\usepackage{amsthm}
\usepackage{amsmath}
\usepackage{stfloats}
\usepackage{amsfonts,amssymb}   
\usepackage[utf8]{inputenc}
\usepackage{booktabs}
\usepackage{graphicx}
\usepackage{epstopdf}
\usepackage{cite}
\usepackage{multirow}
\usepackage[T1]{fontenc}
\usepackage{caption}
\usepackage{float}            
\usepackage{subfloat}
\usepackage{setspace}
\usepackage{color}
\allowdisplaybreaks
\usepackage{easyReview} 

\begin{document}
	
\title{Reconfigurable Intelligent Surface-Enabled Array Radar for Interference Mitigation}
	
\author{Shengyao Chen, Qi Feng, Longyao Ran, Feng Xi, Zhong Liu
\thanks{
	
	
	The authors are with the School of Electronic and Optical Engineering, Nanjing University of Science and Technology, Nanjing 210094, China (e-mail:  chenshengyao@njust.edu.cn, qfeng@njust.edu.cn, rly@njust.edu.cn, xifeng@njust.edu.cn, eezliu@njust.edu.cn).}
}
	
\maketitle
	
\begin{abstract}
Conventional active array radars often jointly design the transmit and receive beamforming for effectively suppressing interferences. 
To further promote the interference suppression performance, this paper introduces a reconfigurable intelligent surface (RIS) to assist the radar receiver because the RIS has the ability to bring plentiful additional degrees-of-freedom.
To maximize the output signal-to-interference-plus-noise ratio (SINR) of receive array, we formulate the codesign of transmit beamforming and RIS-assisted receive beamforming into a nonconvex constrained fractional programming problem, and then propose an alternating minimization-based algorithm to jointly optimize the transmit beamformer, receive beamformer and RIS reflection coefficients. 
Concretely, we translate the RIS reflection coefficients design into a series of unimodular quadratic programming (UQP) subproblems by employing the Dinkelbach transform, and offer the closed-form optimal solutions of transmit and receive beamformers according to the minimum variance distortionless response principle.
To tackle the UQP subproblems efficiently, we propose a second-order Riemannian Newton method (RNM) with improved Riemannian Newton direction, which avoids the line search and has better convergence speed than typical first-order Riemannian manifold optimization methods.
Moreover, we derive the convergence of the proposed codesign algorithm by deducing the explicit convergence condition of RNM.
We also analyze the computational complexity. 
Numerical results demonstrate that the proposed RIS-assisted array radar has superior performance of interference suppression to the RIS-free one, and the SINR improvement is proportional to the number of RIS elements.

\end{abstract}
	
	
\IEEEpeerreviewmaketitle
	

\section{Introduction}
Reconfigurable intelligent surface (RIS) has emerged as a promising solution to improve the performance and extend the wireless signal coverage via passive signal reflection.
It is conventionally a planar array containing a large number of meta-surface reflecting elements and has the capability of adjusting the phases of incident signals in a programmable manner.
Owing to its low cost, high power efficiency and flexible deployment, the RIS has recently aroused growing interest and been envisioned as a fundamental enabling technique for the future wireless communications and networks \cite{Wu2020Towards,Mei2022Intelligent}.

Thanks to deploying RISs, traditional wireless communication systems harvest plenty of additional degrees-of-freedom (DoFs) to drastically improve their performance.
Since RISs have the ability to adjust wireless channels by intelligently controlling the signal reflection, it paves the way to realize an intelligent and programmable wireless environment and thus promotes the paradigm shift of wireless communications from accommodating channels to customizing channels. 
In recent years, RISs are widely deployed in wireless communication systems to transfer passive information \cite{Yan2019Passive}, achieve secure physical-layer transmission \cite{Cui2019Secure, Dong2020Enhancing, Alexandropoulos2021Safeguarding}, and conduct index modulation scheme \cite{Basar2020Reconfigurable}, in virtue of the flexibility of phase shift adjustment.
Furthermore, it has been proved that a large-scale RIS can generate a considerable signal-to-noise ratio (SNR) gain over the cascaded channel \cite{Wu2019Beamforming}.
In light of this fact, RISs are also extensively used to combat the unfavourable channel environments by producing desired passive beams \cite{Rahal2022Arbitrary}, so as to enhance the achievable rate \cite{Huang2018Achievable}, outage probability \cite{Guo2020Outage}, spectral/energy efficiency \cite{Huang2019Reconfigurable} and bit-error-rate performance \cite{Li2021Passive} of wireless communication systems. 

Inspired by significant advantages emerged in wireless communications, equipping radars with RISs is naturally taken into consideration \cite{Liu2023Integrated}.
In its infancy, RISs are utilized for non-line-of-sight (NLoS) target detection because it has the ability to create a virtual path between the radar and the potential target in shadowed areas \cite{Aubry2021Reconfigurable, Song2022Intelligent, Esmaeilbeig2022IRS}.
Compared to conventional NLoS radars \cite{Tang2022Multipath,Chen2022Joint}, RIS-assisted NLoS radars have no requirement on the geometric prior knowledge of shadowed areas, and their signal processing strategy is almost identical to traditional line-of-sight (LoS) radars. 
When the target is not located in shadowed areas, the deployed RIS can also be employed to provide an additional path to strengthen target echoes \cite{Wang2021Joint,Buzzi2021Radar,Buzzi2022Foundations,Zhang2022Metaradar,Lu2021Target}. 
In \cite{Wang2021Joint}, the authors deployed two RISs around the monostatic radar to alleviate the target blockage in cluttered environments. 
In \cite{Buzzi2021Radar}, Buzzi equipped the monostatic radar with a closely-spaced RIS to assist the radar transmitter or receiver for improving the target detection performance. 
This preponderance is then extended to bistatic \cite{Buzzi2022Foundations}, colocated MIMO \cite{Zhang2022Metaradar} and distributed MIMO radars \cite{Lu2021Target}, respectively.

Additional DoFs introduced by RISs into radar systems are also applicable to enhance the signal processing performance \cite{Shao2022Target,Esmaeilbeig2022Cramer,Esmaeilbeig2023Joint,Liu2023Joint}. 
To improve the direction-of-arrival (DoA) estimation accuracy, Shao et al. deployed a RIS into a colocated array radar for maximizing the minimum receive SNRs of all potential target directions in \cite{Shao2022Target}.
Esmaeilbeig et al. equipped the pulse-Doppler radar with several RISs in \cite{Esmaeilbeig2022Cramer} to improve the performance of target parameter estimation via the Cramer-Rao lower bound minimization.  
This RIS configuration strategy is thereafter extended to the colocated MIMO radar with the aid of transmit waveform design \cite{Esmaeilbeig2023Joint}.
Moreover, to boost the ability of multi-target detection, Liu et al. jointly designed the transmit waveform, RIS reflection coefficients and receive filter of multiple RIS-assisted colocated MIMO radar for improving the interference suppression performance \cite{Liu2023Joint}.



Owing to the ability to offer extra DoFs, RISs are naturally a promising approach enabling array radar systems to promote the interference suppression performance.
To reduce the effect of ground clutter, we have equipped an active array radar with a RIS to decrease the sidelobe level in transmit beampattern synthesis \cite{Ran2023Beampattern}. 
However, this scheme has a limited ability of interference suppression and thus will be difficult to achieve the requirements of weak target detection, such as pedestrian or low-altitude unmanned aerial vehicle (UAV) detection in applications of vehicular and security radars.
This paper therefore devotes to providing a powerful interference suppression scheme for active array radars with the assistance of RISs.

In traditional active array radars, interference suppression is usually carried out by resorting to the codesign of transmit and receive beamforming \cite{Liu2014Joint,Chen2014Adaptive}. 
However, its performance mainly depends on the DoFs of transmit and receive arrays.
This paper employs an RIS to assist the receive array and thus provides a great number of additional DoFs because the RIS often has significantly more elements than the receive array.
Consequently, the RIS-assisted active array radar has the potential of suppressing more interferences than the traditional one.
Moreover, the passive beamforming of RIS has the ability to yield a large processing gain via the phase alignment of LoS and NLoS paths \cite{Buzzi2021Radar}, which is applicable to improve the output signal-to-interference-plus-noise ratio (SINR) of the proposed RIS-assisted array radar.

To maximize the output SINR of receive array, this paper dedicates to jointly designing the active transmit and receive beamformers and the passive RIS beamformer under the total power constraint of transmit beamformer and the unimodular constraints of RIS reflection coefficients.
The main contributions are summarized as follows:

\begin{itemize}
\item \emph{Providing an effective beamforming codesign algorithm for RIS-assisted active array radar: }
To maximize the output SINR, we coin the proposed beamforming codesign as a nonconvex constrained fractional programming (FP) problem and optimize the transmit beamformer, receiver beamformer and RIS reflection coefficients via alternating minimization.
Specifically, we consider the optimization of RIS reflection coefficients as a unimodular constrained FP (UFP) problem and provide a customized iterative solution algorithm by exploiting the Dinkelbach transform and the proposed Riemannian Newton method (RNM).
We also give the closed-form optimal solutions of both transmit and receive beamformers according to the minimum variance distortionless response (MVDR) principle. 
Thanks to the solving methods of all three beamformers, the proposed beamforming codesign algorithm is effective and computationally efficient.

\item \emph{Proposing an efficient Riemannian Newton method for solving unimodular quadratic programming:} 
For the optimization of RIS reflection coefficients, we translate the UFP into a series of unimodulus quadratic programming (UQP) subproblems by using the Dinkelbach transform.
Then we propose an efficient second-order Riemannian manifold optimization (RMO) approach, RNM, to tackle all UQP subproblems.
In the proposed RNM, we explicitly give the Riemannian Hessian matrix of the objective function and then introduce a diagonal loading matrix into the Riemannian Newton equation to ensure that the improved Riemannian Newton direction is always descent.
Owing to the faster convergent speed and avoidance of line search, the RNM consumes less computational cost than typical first-order RMO methods.

\item \emph{Analyzing the convergence and computational complexity:}
We first establish the explicit convergence condition of the proposed RNM. 
To the best of our knowledge, this is the first time about the second-order RMO approach without line search. 
Based on this condition, we prove that the output SINR is upper bounded and monotonically non-decreasing in each iteration of alternating minimization, and therefore necessarily converges to a finite value. 
We also analyze the computational complexity of the proposed beamforming codesign algorithm, which is mainly dominated by the step of the RIS optimization.

\item \emph{Conducting extensive numerical experiments:} 
Experiment results demonstrate that deploying a closely-spaced RIS can remarkably promote the performance of interference suppression for active array radars and the output SINR improvement is basically proportional to the RIS element number.
Moreover, the assistance of RIS offers the ability to reduce the requirement on the number of radar array elements without the SINR performance loss, providing a promising solution to enhance the flexibility of active array radar design and to reduce the cost and complexity. 

\end{itemize}

It is worth pointing out that we have preliminarily verified that deploying an active RIS considerably improves the SINR performance of array radar in \cite{Feng2023Joint}.
Compared to widely-used passive RISs, the active RIS can offer more DoFs since it has the ability to amplify incident signals \cite{Zhang2023ActiveRIS}.
But the expense is more power consumption and hardware complexity, as well as the introduction of amplified incident noise. 
Moreover, the reflection coefficients of active RIS are limited by the total power and the maximum amplification factor simultaneously, and thus are difficult to solve in large-scale RIS scenes.

The rest of this paper is organized as follows. 
The signal model and problem formulation is presented in Section II. 
The alternating minimization-based codesign algorithm is proposed to optimize the RIS reflection coefficients, receive beamformer and transmit beamformer in Section III. 
Section IV analyzes the convergence and computational complexity of the proposed codesign algorithm. 
Section V verifies the superior performance of the RIS-assisted array radar and the effectiveness of the proposed codesign algorithm via several experiments.
Section VI finally draws some conclusions and~perspectives. 

$\emph{Notations}$: 
Vectors and matrices are denoted by boldface lowercase and uppercase letters, respectively. 
The $(\cdot)^{T}$ , $(\cdot)^{H}$, $(\cdot)^{*}$ and $|\cdot|$ are the transpose, conjugate transpose, complex conjugate and modulus operators. 
$\mathrm{Re}\{\cdot\}$ and $\mathrm{Im}\{\cdot\}$ denote the real and imaginary components of a complex number, respectively.
$\bm{0}_{M}$, $\bm{1}_{M}$, $\bm{O}_M$ and $\bm{I}_M$ represent the $M\times 1$ all-zero vector,  $M\times 1$ all-one vector, $M\times M$ all-zero matrix and $M\times M$ identity matrix.
$\mathrm{diag}(\bm{X})$ is the vector formulated by the diagonal elements of $\bm{X}$, and $\mathrm{Diag}(\bm{x})$ is the diagonal matrix with the elements of $\bm{x}$ on the diagonal.
$\otimes$ and $\odot$ stand for Kronecker product and Hadamard product, respectively.
$\Vert\cdot\Vert_2$ and $\Vert\cdot\Vert_{\text F}$ stand for Euclidean norm and Frobenius norm.

\begin{figure}[t!]
	\centering
	\includegraphics[width=0.5\linewidth]{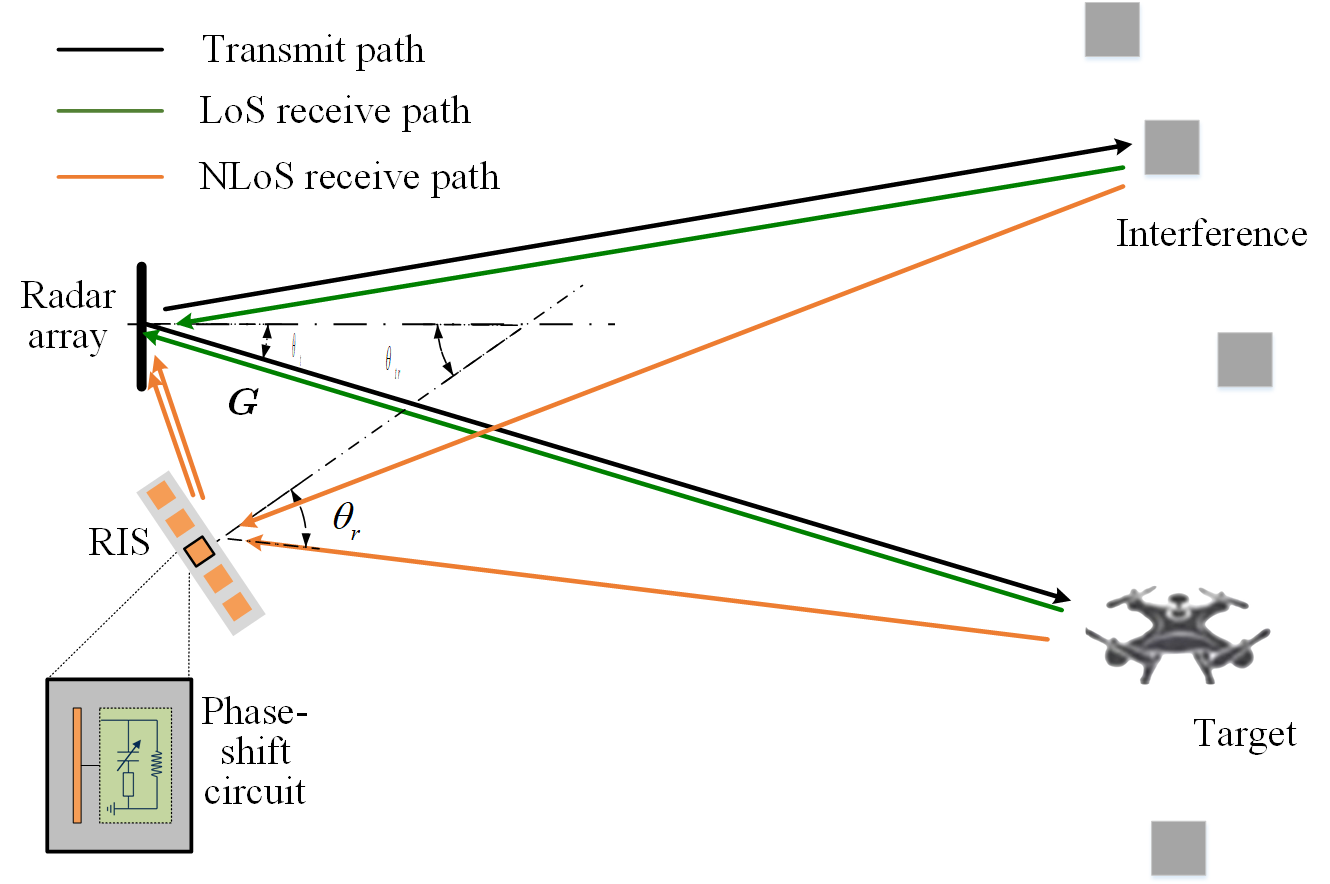}
	\caption{RIS-assisted active array radar}
	\label{fig:systemlayout}
	\vspace{-6mm}
\end{figure}

\section{Signal Model and Problem Formulation}
Let us consider a RIS-assisted array radar which consists of an active array transceiver and an RIS, as shown in Fig. 1.
The RIS is a uniform linear array (ULA) containing $M$ metasurface reflecting elements, and the radar transmit and receive arrays share an $N$-element ULA.
Assume the transmit signal satisfying the narrowband criterion and the desired target meeting the far-field point target hypothesis.
The target angle-of-arrival with respect to the radar array, denoted as $\theta_t$, and that with respect to the RIS, denoted as $\theta_r$, are clearly different. 
They often follow the relationship $\theta_r=\theta_t+\theta_{tr}$, where $\theta_{tr}$ represents the angle between the radar array and the RIS \cite{Esmaeilbeig2023Joint}.
As seen in Fig. 1, the target and interference echoes are received by the RIS-assisted array radar receiver. 
That is to say, the target and interference echoes pass through two paths, the LoS path (target/interference-radar) and the NLoS path (target/interference-RIS-radar).

Denote $s(l)$ as the transmit waveform of array radar, $l=1,2,\cdots,L$, with the number of samples $L$ in a transmit pulse, and $\bm{u}\triangleq [u_1,u_2,\cdots ,u_{N}]^T\in\mathbb{C}^{N}$ as the transmit beamformer weights.
Generally, the transmit beamformer $\bm{u}$ is limited by the total radar transmit power. 
For simplicity, here we set
\begin{equation}
\label{Radarpower}
\Vert \bm{u} \Vert_2^2=1.
\end{equation}

For a target at the look direction $\theta$ from the active array, the achieved signal at the target location is expressed as
\begin{equation}
\label{ylos}
    x_{\rm{T}}(l)=\bm{a}^T(\theta)\bm{u}s(l),
\end{equation}
in which $\bm{a}(\theta)$ is the active array steering vector given by
\begin{equation}
\label{RSvector}
    \bm{a}(\theta) =[1,e^{-j\nu d_{\rm R}\sin\theta},\cdots,e^{-j\nu (N-1)d_{\rm R} \sin\theta}]^T,
\end{equation}
where $\nu$ is the wavenumber of transmit waveform and $d_{\rm R}$ is the element spacing of active array.
Note that the target scattering echo reaches the radar receiver through two propagation paths.
The target echo arriving to the receive array is given by
\begin{equation}
	\begin{aligned}
		\label{rtsignal}
		\bm{x}_{\rm{R}}(l) &=\bm{a}(\theta){x}_{\rm{T}}(l)+\bm{G}^T\bm{V}\bm{b}(\theta+\theta_{tr}){x}_{\rm{T}}(l) \\
		&=(\bm{a}(\theta)+\bm{G}^T\bm{V}\bm{b}(\theta+\theta_{tr}))\bm{a}^T(\theta)\bm{u}s(l)
	\end{aligned}
\end{equation}
where $\bm{b}(\theta)$ is the RIS steering vector given by
\begin{equation}
	\label{RISSvector}
	\bm{b}(\theta) =[1,e^{-j\nu d_{\rm I}\sin\theta}, \cdots, e^{-j\nu (M-1)d_{\rm I} \sin\theta}]^T,
\end{equation}
with the RIS element spacing $d_{\rm I}$; 
$\bm{G}$ stands for the channel matrix between the active array and the RIS; 
$\bm{V} \triangleq \mathrm{Diag}([v_{1},v_{2},\cdots,v_{M}])$ is the RIS reflection coefficient matrix with $v_m=p_m e^{j\phi_m}$, where $p_m \in [0,1]$ and $\phi_m \in [0,2\pi]$ respectively denote the amplitude reflection coefficient and the phase shift of the $m$-th element for $m=1,\cdots M$. 
It is commonly set $p_m=1$ to obtain the maximum reflection power, namely, the RIS reflection coefficients are constrained~by
\begin{equation}
|v_m| = 1, \ \forall m, 
\end{equation}
and $\bm{V}$ is often expressed by its diagonal vector $\bm{v} \triangleq \mathrm{diag}({\bm V})$.

Consider that a potential target appears at the look direction $\theta_0$ and $I$ interferences respectively appear at the look direction $\theta_i$, $i=1,2,\cdots,I$, from the radar array with $\theta_i \neq \theta_0$.
Denote $\tau_0$ and $\tau_i$ as the time-delays of the target and the $i$-th interference, respectively, and all the interferences are located at the range cells near the potential target.
Then the total echo received by the active array is given by
\begin{equation}
	\label{rsignal}
	\bm{x}_{\rm{total}}(l)=\tilde{\gamma}_0 \bm{\Phi}(\theta_0)\bm{u}s(l-\tau_{0}) +\sum_{i=1}^{I}\tilde{\gamma}_i\bm{\Phi}(\theta_i)\bm{u}s(l-\tau_{i}) +\bm{n}_0(l),
\end{equation}
in which $\tilde{\gamma}_0$ and $\tilde{\gamma}_i$ are respectively the complex scattering coefficients of potential target and $i$-th interference, $i=1,2,\cdots,I$, $\bm{n}_0(l)$ is the Gaussian noise vector, and
\begin{equation}
\label{eq:Psitheta}
    \bm{\Phi}(\theta)=(\bm{a}(\theta)+\bm{G}^T\bm{V}\bm{b}(\theta+\theta_{tr})) \bm{a}^T(\theta).
\end{equation}

Let us process the received echo $\bm{x}_{\rm{total}}(l)$ through matched filtering and then align the output to the range cell of potential target.
Then the aligned output is given by
\begin{equation}
    \bm{x}_{\rm{MF}} =\gamma_0 \bm{\Phi}(\theta_0)\bm{u}+\sum_{i=1}^{I}\gamma_i \bm{\Phi}(\theta_i)\bm{u}+\bm{n},
\end{equation}
in which ${\gamma}_0$ and ${\gamma}_i$ are respectively the complex scattering coefficients of potential target and $i$-th interference after the process of matched filter, and $\bm{n}$ is the Gaussian noise vector whose distribution satisfies $\bm{n}\sim\mathcal{N}(0,\sigma_n^2\bm{I}_N)$.
It is usually assumed that all $\gamma_i$ are statistically independent and have zero mean and variance $\sigma_i^2$ \cite{Liu2014Joint}. 

The signal $\bm{x}_{\rm{MF}}$ is then spatially filtered by the receive beamformer $\bm{w}$, and the obtained output SINR is given by
\begin{equation}
\label{SINR}
    \text{SINR}=\frac{\vert\gamma_0\vert^2|\bm{w}^H \bm{\Phi}(\theta_0)\bm{u}|^2}{\bm{w}^H\bm{\Psi_u}\bm{w}+\sigma_n^2\bm{w}^H\bm{w}},
\end{equation}
where $\bm{\Psi_u}=\sum_{i=1}^{I}\sigma_i^2 \bm{\Phi}(\theta_i)\bm{u}\bm{u}^H\bm{\Phi}^H(\theta_i)$.
As we known, the performance of target detection often dramatically relies on the output SINR.
Therefore, this paper tries to optimize the output SINR through the joint design of transmit beamformer $\bm{u}$, RIS reflection coefficients $\bm{v}$ and receive beamformer $\bm{w}$.

According to aforementioned analysis, the proposed codesign of transmit and RIS-assisted receive beamforming can be formulated as
\begin{subequations}
\label{Optprob1}
	\begin{align}
	\label{Optprob11}
    \max_{\bm{u},\bm{v},\bm{w}} & \ G(\bm{u},\bm{v},\bm{w}) \triangleq \frac{|\bm{w}^H \bm{\Phi}(\theta_0)\bm{u}|^2} {\bm{w}^H\bm{\Psi_u}\bm{w}+\sigma_n^2\bm{w}^H\bm{w}}   \\
    \label{Optprob12}
    \text{s.t.} \ & \ \Vert \bm{u} \Vert_2^2=1 \\
    \label{Optprob13}
    & \ |v_{m}|=1, \ \forall m.
\end{align}
\end{subequations}
Note that the constraints \eqref{Optprob12} and \eqref{Optprob13} are both nonconvex, and the objective function $G(\bm{u},\bm{v},\bm{w})$ is a nonconvex fractional function.
It is intractable to tackle the problem \eqref{Optprob1} directly by exploiting existing methods.
In the following, we construct an alternating minimization-based method to solve \eqref{Optprob1} efficiently with the assistance of FP and RMO techniques.

\section{The Proposed Beamforming Codesign Algorithm \label{sec:alg}}
By using the alternating minimization scheme, we decompose the problem \eqref{Optprob1} into several subproblems and then separately solve the variables $\bm{v}$, $\bm{w}$ and $\bm{u}$ in \eqref{Optprob1} in virtue of the Dinkelbach transform, second-order RMO and MVDR.

\subsection{Optimization of Passive RIS Beamforming}
Fixing the variables $\bm{u}$, $\bm{w}$ and ignoring irrelative constant items, we can rewrite \eqref{Optprob1} with respect to $\bm{v}$ as
\begin{subequations}
	\label{Optprob_v}
	\begin{align}
		\label{Optprob_vobj}
		&\max_{\bm{v}} \ E(\bm{v})=\frac{|\bm{w}^H \bm{\Phi}(\theta_0)\bm{u}|^2}
		{\sum_{i=1}^{I}\sigma_i^2|\bm{w}^H \bm{\Phi}(\theta_i)\bm{u}|^2+\sigma_n^2\bm{w}^H\bm{w}}   \\
		\label{Optprob_vcon}
		&\ \text{s.t.} \quad |v_{m}|=1, \ \forall m.
	\end{align}
\end{subequations}	
Before proceeding, we explicitly represent the objection function $E(\bm{v})$ with respect to the variable $\bm{v}$.
Note that the function $\bm{w}^H \bm{\Phi}(\theta)\bm{u}$ with respect to $\bm{v}$ can be described as 
\begin{equation}
	\label{wPsit}
	\begin{split}
		\bm{w}^H \bm{\Phi}(\theta)\bm{u} 
		&=\bm{w}^H\bm{a}(\theta)\bm{a}^T(\theta)\bm{u} +\bm{w}^H\bm{G}^T\bm{V}\bm{b}(\theta+\theta_{tr})\bm{a}^T(\theta)\bm{u}  \\
		&=\bm{h}^H(\theta)\bm{v}+\beta(\theta),  \\
	\end{split}
\end{equation}
where $\bm{h}(\theta)=\bm{a}^{H}(\theta)\bm{u}^{*}\mathrm{diag}(\bm{b}^{*}(\theta+\theta_{tr}))\bm{G}^{*}\bm{w}$ and $\beta(\theta)=\bm{w}^H\bm{a}(\theta)\bm{a}^T(\theta)\bm{u}$.
Inserting \eqref{wPsit} into \eqref{Optprob_v}, we rewrite \eqref{Optprob_v} as
\begin{equation}
	\label{Optprob_v1}
	\begin{split}
		&\max_{\bm{v}} \ \frac{|\bm{h}^H(\theta_0)\bm{v}+\beta(\theta_0)|^2}
		{\sum_{i=1}^{I}\sigma_i^2|\bm{h}^H(\theta_i)\bm{v}+\beta(\theta_i)|^2+\sigma_n^2\bm{w}^H\bm{w}} \\
		&\ \text{s.t.} \quad |v_{m}|=1, \ \forall m.  \\
	\end{split}
\end{equation}

We observe that \eqref{Optprob_v1} is a nonconvex constrained FP problem.
In virtue of the Dinkelbach transform \cite{Shen2018Fractional}, we can translate the fractional objective function of \eqref{Optprob_v1} into a quadratic objective function, and then express the problem \eqref{Optprob_v1} as a series of subproblems as follows, 
\begin{equation}
	\label{Dinkelbach_v1}
	\begin{split}
		\min_{\bm{v}} &\ \sum_{i=1}^{I}z\sigma_i^2|\bm{h}^H(\theta_i)\bm{v}+\beta(\theta_i)|^2 +z\sigma_n^2\bm{w}^H\bm{w} \\
		& \ -|\bm{h}^H(\theta_0)\bm{v}+\beta(\theta_0)|^2   \\
		\text{s.t.} & \  |v_{m}|=1, \ \forall m,  \\
	\end{split}
\end{equation}
in which $z$ is the auxiliary variable and its value is given by
\begin{equation}
	\label{Dinkpar_v}
	z^{(q)}=\left. \frac{|\bm{h}^H(\theta_0)\bm{v}+\beta(\theta_0)|^2}
	{\sum_{i=1}^{I}\sigma_i^2|\bm{h}^H(\theta_i)\bm{v}+\beta(\theta_i)|^2+\sigma_n^2\bm{w}^H\bm{w}} \right|_{\bm{v}=\bm{v}^{(q)}},
\end{equation}
at the $q$-th iteration. Removing constant terms, we rewrite the subproblem \eqref{Dinkelbach_v1} as
\begin{equation}
\label{Dinkelbach_v2}
\begin{split}
\min_{\bm{v}} &\ f^{\prime}(\bm{v}) \triangleq \bm{v}^H(\sum_{i=1}^{I}z\sigma_i^2\bm{H}_i-\bm{H}_0)\bm{v} +2\mathrm{Re}\{\bm{v}^H\tilde{\bm{h}}(z)\} \\
\text{s.t.} & \ |v_{m}|=1, \ \forall m,  \\
\end{split}
\end{equation}
where $\tilde{\bm{h}}(z)=\sum_{i=1}^{I}z\sigma_i^2\beta(\theta_i)\bm{h}(\theta_i)-\beta(\theta_0)\bm{h}(\theta_0)$ and $\bm{H}_i=\bm{h}(\theta_i)\bm{h}^H(\theta_i)$ with $i=0,1,\cdots,I$.

It is noted that the problem \eqref{Dinkelbach_v2} is a quadratic programming with unimodular constraints.
Since a unimodular constraint is equivalent to a complex circular manifold (CCM) \cite{Absil2008Optimization}, we can consider \eqref{Dinkelbach_v2} as a unconstrained problem on the CCM.
As a result, we handle the problem \eqref{Dinkelbach_v2} by employing the proposed second-order RNM which has the capability of avoiding line search and has better convergence rate than typical first-order RMO approaches, such as Riemannian gradient descend (RGD) and Riemannian conjugate gradient (RCG).
The basic idea of RNM is first to update the variable along the improved Riemannian Newton direction (RND), which ensures that the search direction is always descent, and thereafter to project the updated variable onto the manifold to guarantee that all the searched variables are necessarily on the manifold.

To ensure that the objective function in \eqref{Dinkelbach_v2} is positive definite, we introduce a regularization term $\lambda_v \bm{v}^H\bm{v}$ into $f^{\prime}(\bm{v})$ with a positive constant $\lambda_v>0$.
Since $\lambda_v \bm{v}^H\bm{v}$ equals the constant $\lambda_v M$ due to unimodular constraints, it has no effect on the solution of \eqref{Dinkelbach_v2}. 
Therefore, \eqref{Dinkelbach_v2} is equivalently expressed as
\begin{subequations}
	\label{Dinkelbach_v3}
	\begin{align}
		\label{Dinkelbach_v3obj}
	    \min_{\bm{v}} &\ f(\bm{v}) \triangleq \bm{v}^H(\sum_{i=1}^{I}z\sigma_i^2\bm{H}_i-\bm{H}_0+\lambda_v\bm{I}_M)\bm{v} \\ 
        & \qquad \quad \ +2\mathrm{Re}\{\bm{v}^H\tilde{\bm{h}}(z)\} \notag \\
		\label{Dinkelbach_v3con}
		&\ \text{s.t.} \quad |v_{m}|=1, \ \forall m,
	\end{align}
\end{subequations}
where $\lambda_v$ satisfies $\lambda_v > \Vert\bm{h}(\theta_0)\Vert_2^2$ to ensure that the matrix $\tilde{\bm{H}}=\sum_{i=1}^{I}z\sigma_i^2\bm{H}_i-\bm{H}_0+\lambda_v\bm{I}_M$ is always positive definite.		

Define the Euclidean gradient of $f(\bm{v})$ with respect to the complex variable $\bm{v} $ as
\begin{equation}
\label{EGradDef}
\triangledown f(\bm{v})= \triangledown_{\bm{v}^{*}} f(\bm{v}),
\end{equation}
where $\triangledown_{\bm{v}^{*}} f(\bm{v})$ denotes the gradient of $f(\bm{v})$ over the variable $\bm{v}^{*}$ \cite{Brandwood1983Complex}.
The Euclidean gradient of $f(\bm{v})$ at $\bm{v}$ is then given by
\begin{equation}
\label{EGradf}
\triangledown f(\bm{v})= \tilde{\bm{H}}\bm{v}+\tilde{\bm{h}}(z),
\end{equation}
and the corresponding Euclidean Hessian is given by
\begin{equation}
\label{EHessf}
\triangledown^2 f(\bm{v})= \tilde{\bm{H}}.
\end{equation}

Define the constraint \eqref{Dinkelbach_v3con} as the CCM $\mathcal{M}\triangleq\{\bm{v}\in \mathbb{C}^M:\bm{v}\odot \bm{v}^{*}=\bm{1}_M\}$ and the corresponding tangent space at $\bm{v}$ as $\mathbb{T}_{\bm v}\mathcal{M}\triangleq\{\bm{q}\in \mathbb{C}^M:\mathrm{Re}\{\bm{q}\odot\bm{v}^*\}=\bm{0}_M\}$ \cite{Alhujaili2019Transmit}.
By projecting the Euclidean gradient $\triangledown f(\bm{v})$ onto the tangent space $\mathbb{T}_{\bm v}\mathcal{M}$, we obtain the Riemannian gradient as
\begin{equation}
\label{RiemGradf}
   \mathrm{grad} f(\bm{v})=\mathcal{P}_{\bm{v}}\triangledown f(\bm{v}) =\triangledown f(\bm{v}) -\mathrm{Re}\{\triangledown f(\bm{v}) \odot \bm{v}^{*}\} \odot \bm{v},
\end{equation}
where $\mathcal{P}_{\bm{v}}(\bm{y})=\bm{y}-\mathrm{Re}\{\bm{y}\odot \bm{v}^{*}\} \odot \bm{v}$ is orthogonal projector onto $\mathbb{T}_{\bm{v}}\mathcal{M}$.
Moreover, we obtain the Riemannian Hessian of $f(\bm{v})$ at the point $\bm{v}$ as
\begin{equation}
\begin{split}
	\label{RiemHessf1}
    \mathrm{Hess} f(\bm{v})[\bm{\xi}] =\mathcal{P}_{\bm{v}}\triangledown^2 f(\bm{v})\bm{\xi} +\mathfrak{A}_{\bm{v}}(\bm{\xi},\mathcal{P}_{\bm{v}}^{\perp}\triangledown f(\bm{v})),
\end{split}
\end{equation}
for all $\bm{\xi} \in \mathbb{T}_{\bm{v}}\mathcal{M}$, where $\mathfrak{A}_{\bm{v}}(\cdot,\cdot)$ is the Weingarten map operator of the manifold $\mathcal{M}$ at $\bm{v}$ \cite{Absil2013Extrinsic}, and $\mathcal{P}_{\bm{v}}^{\perp} =\bm{I}_M-\mathcal{P}_{\bm{v}}$ is the orthogonal projector onto the normal space to $\mathcal{M}$ at $\bm{v}$.

Since $\mathcal{M}$ is an oblique manifold, we have
\begin{equation}
\label{ProjHess}
	\begin{split}
	\mathcal{P}_{\bm{v}}\triangledown^2 f(\bm{v})\bm{\xi} &=\tilde{\bm{H}}\bm{\xi} -\mathrm{Re}\{\tilde{\bm{H}}\bm{\xi} \odot \bm{v}^{*}\} \odot \bm{v}   \\
	&=\tilde{\bm{H}}\bm{\xi} -\mathrm{Diag}(\bm{v})\mathrm{Re}\{\mathrm{Diag}(\bm{v}^{*})\tilde{\bm{H}}\bm{\xi}\}
	\end{split}
\end{equation}
and 
\begin{equation}
\label{WeingRGrad}
	\begin{split}
	\mathfrak{A}_{\bm{v}}(\bm{\xi},\mathcal{P}_{\bm{v}}^{\perp}\triangledown f(\bm{v})) &=-\mathcal{P}_{\bm{v}} \mathrm{Re}\{\triangledown f(\bm{v}) \odot\bm{v}^{*}\} \odot \bm{\xi}  \\	
	&=-\mathrm{Re}\{\triangledown f(\bm{v}) \odot\bm{v}^{*}\} \odot \bm{\xi}  \\
	& \quad +\mathrm{Re}\{(\mathrm{Re}\{\triangledown f(\bm{v}) \odot\bm{v}^{*}\} \odot \bm{\xi})\odot \bm{v}^{*}\} \odot \bm{v}  \\
	&=-\bm{Q}\bm{\xi} +\mathrm{Diag}(\bm{v})\mathrm{Re}\{\mathrm{Diag}(\bm{v}^{*})\bm{Q}\bm{\xi}\}
	\end{split}
\end{equation}
where $\bm{Q}=\mathrm{Diag}(\mathrm{Re}\{\triangledown f(\bm{v}) \odot\bm{v}^{*}\})$ \cite{Hu2020Brief}.
Inserting \eqref{ProjHess} and \eqref{WeingRGrad} into \eqref{RiemHessf1}, we obtain
\begin{equation}
\begin{split}
\label{RiemHessf2}
   \mathrm{Hess} f(\bm{v})[\bm{\xi}] = & (\tilde{\bm{H}}-\bm{Q})\bm{\xi} \\ &-\mathrm{Diag}(\bm{v})\mathrm{Re}\{\mathrm{Diag}(\bm{v}^{*})(\tilde{\bm{H}}-\bm{Q})\bm{\xi}\}
\end{split}
\end{equation}
In the standard Riemannian Newton method, we need to produce the direction $\bm{\xi} \in \mathbb{T}_{\bm{v}}\mathcal{M}$ satisfying
\begin{equation}
	\begin{split}
		\label{NewtowEq}
		\mathrm{grad} f(\bm{v})+ \mathrm{Hess} f(\bm{v})[\bm{\xi}] =0.
	\end{split}
\end{equation}

Denote $\hat{\bm{v}}=[\bm{v}_R^T,\bm{v}_I^T]^T$ with $\bm{v}_R=\mathrm{Re}\{\bm{v}\}$ and $\bm{v}_I=\mathrm{Im}\{\bm{v}\}$, 
$\hat{\bm{g}}=[\bm{g}_R^T,\bm{g}_I^T]^T$ with $\bm{g}_R=\mathrm{Re}\{\mathrm{grad} f(\bm{v})\}$ and $\bm{g}_I=\mathrm{Im}\{\mathrm{grad} f(\bm{v})\}$,
$\hat{\bm{\xi}}=[\bm{\xi}_R^T,\bm{\xi}_I^T]^T$ with $\bm{\xi}_R=\mathrm{Re}\{\bm{\xi}\}$ and $\bm{\xi}_I=\mathrm{Im}\{\bm{\xi}\}$, 
and express the matrices $\tilde{\bm{H}}$ and $\bm{Q}$, respectively, by their equivalent real matrices
\begin{equation}	
\hat{\bm{H}}=\left[             \notag
\begin{array}{cc}   
\tilde{\bm{H}}_R & -\tilde{\bm{H}}_I \\ 
\tilde{\bm{H}}_I & \tilde{\bm{H}}_R \\
\end{array}
\right]    \ \textrm{and}  \  \       
\hat{\bm{Q}}=\left[             \notag
\begin{array}{cc}   
	\bm{Q} & \  \bm{O}_M \\ 
	\ \bm{O}_M & \bm{Q} \\
\end{array} 
\right],
\end{equation}
where $\tilde{\bm{H}}_R=\mathrm{Re}\{\tilde{\bm{H}}\}$ and $\tilde{\bm{H}}_I=\mathrm{Im}\{\tilde{\bm{H}}\}$.
We convert the Newton equation \eqref{NewtowEq} into its real number form as
\begin{equation}
\label{NewtonEq}
\begin{split}
\hat{\bm{g}} +\frac{1}{2}(\bm{I}_{2M}-\hat{\bm{V}})(\hat{\bm{H}}-\hat{\bm{Q}})\hat{\bm{\xi}}=0,
\end{split}
\end{equation}
where 
\begin{equation}	
\hat{\bm{V}}=\left[             \notag
\begin{array}{cc}   
\mathrm{Re}\{\mathrm{Diag}(\bm{v}\odot\bm{v})\} & \mathrm{Im}\{\mathrm{Diag}(\bm{v}\odot\bm{v})\} \\ 
\mathrm{Im}\{\mathrm{Diag}(\bm{v}\odot\bm{v})\} & -\mathrm{Re}\{\mathrm{Diag}(\bm{v}\odot\bm{v})\} \\
\end{array}
\right].               
\end{equation}

Define the real Riemannian Hessian matrix as
\begin{equation}
\label{HessR}
\mathrm{Hess_{R}} f(\bm{v})=\frac{1}{2}(\bm{I}_{2M}-\hat{\bm{V}})(\hat{\bm{H}}-\hat{\bm{Q}})
\end{equation}
We find that $\mathrm{Hess_{R}} f(\bm{v})$ is not necessarily positive definite, which cannot ensure that the solution of \eqref{NewtonEq} is a descent direction of $f(\bm{v})$.
To generate a descent Newton direction, we replace \eqref{NewtonEq} with the following equation, 
\begin{equation}
	\label{NewtonEq1}
	\begin{split}
		\hat{\bm{g}} +(\mathrm{Hess_{R}} f(\bm{v})+\bm{E})\hat{\bm{\xi}}=0,
	\end{split}
\end{equation}
where $\bm{E}$ is the regularization matrix to make $\mathrm{Hess_{R}} f(\bm{v})+\bm{E}$ positive definite \cite{Absil2008Optimization}.
For the convenience of implementation, we set $\bm{E}=\frac{\mu}{2}\bm{I}_{2M}$, where ${\mu}>0$ is the diagonal loading factor.
The equation \eqref{NewtonEq1} is therefore converted into
\begin{equation}
	\label{NewtonEq2}
	\begin{split}
		\hat{\bm{g}} +(\mathrm{Hess_{R}} f(\bm{v})+\frac{\mu}{2}\bm{I}_{2M})\hat{\bm{\xi}}=0,
	\end{split}
\end{equation}

For a suitable choice of $\mu$, the solution of \eqref{NewtonEq2} is always a descent direction.
It is worth noting that the solution of \eqref{NewtonEq2} is the standard RND when $\mu=0$ and becomes the Riemannian gradient when $\mu=\infty$. 
In fact, the proposed improved RND makes a compromise between the standard RND and the Riemannian gradient to guarantee the decrease of objection function $f(\bm{v})$.
Then, we obtain the improved RND~as 
\begin{equation}
\begin{split}
   \label{NewtowDirect}
   \hat{\bm{\xi}}^{\star} = -(\mathrm{Hess_{R}} f(\bm{v})+\frac{\mu}{2}\bm{I}_{2M})^{-1}\hat{\bm{g}},
\end{split}
\end{equation}  

\noindent and its complex number form as 
\begin{equation}
   \label{CNewtowDirect}
   \bm{\xi}^{\star}=\bm{\xi}_R^{\star}+j\bm{\xi}_I^{\star}.
\end{equation}

\begin{algorithm}[t!]
	\label{alg:RMO}
	\caption{RNM-based Algorithm for Solving \eqref{Optprob_v}.}
	\begin{algorithmic}[1]
		\renewcommand{\algorithmicrequire}{\textbf{Input:}}
		\REQUIRE $\bm{w}^{p}$, $\bm{u}^{p}$, $\beta(\theta_i)$, $\bm{h}(\theta_i)$ and $\sigma_i^2$. \\        
		\renewcommand{\algorithmicensure}{\textbf{Output:}}
		\ENSURE RIS reflection coefficients $\bm{v}^{p+1}$.  
		\STATE Set $q=0$ and initialize $\bm{v}^{(0)}=\bm{v}^{p}$.   
		\REPEAT
		\STATE Set $\check{\bm{v}}^{(0)}=\bm{v}^{(q)}$;
		\STATE Update $z^{(q)}$ using \eqref{Dinkpar_v};		
		\REPEAT
		\STATE Compute the complex RND using \eqref{CNewtowDirect};
		\STATE Obtain $\check{\bm{v}}^{(r+1)}$ using \eqref{RiemaNewUpdate} and \eqref{RiemaProj};
		\STATE $r \leftarrow r+1$;
		\UNTIL{convergence}
		
		\STATE Set $\bm{v}^{(q+1)}=\check{\bm{v}}^{(r)}$;
		\STATE $q \leftarrow q+1$;
		\UNTIL{convergence}
	\end{algorithmic}
    \vspace{-1mm}
\end{algorithm}

Denote $\check{\bm{v}}^{(r)}$ as the searched point of $r$-th iteration and $\bm{\xi}(\check{\bm{v}}^{(r)})$ as the corresponding complex RND obtained by \eqref{CNewtowDirect}.
Along the direction $\bm{\xi}(\check{\bm{v}}^{(r)})$, we update the searched point at the $(r+1)$-th iteration as
\begin{equation}
   \label{RiemaNewUpdate}
   \tilde{\bm{v}}^{(r+1)}=\check{\bm{v}}^{(r)} +\bm{\xi}(\check{\bm{v}}^{(r)}),
\end{equation}
Thereafter, we map $\tilde{\bm{v}}^{(r+1)}$ onto $\mathcal{M}$ to generate the $(r+1)$-th iterative output as
\begin{align}
\label{RiemaProj}
\check{\bm{v}}^{(r+1)}=\mathcal{R}_{\bm{v}^{(r)}}(\tilde{\bm{v}}^{(r+1)}),
\end{align}
in which
$\mathcal{R}_{\bm{v}}(\cdot)\colon \mathbb{T}_{\bm{v}}\mathcal{M}\to\mathcal{M}$ denotes the retraction operation, which projects the point on the tangent space $\mathbb{T}_{\bm{v}}\mathcal{M}$ to the manifold $\mathcal{M}$. 
An extensively-used scheme is to individually project each element of $\tilde{\bm{v}}^{(r+1)}$ onto a unit circle, namely, 
\begin{align}
\label{eqR}
(\mathcal{R}_{\bm{v}^{(r)}}(\tilde{\bm{v}}^{(r+1)}))_{m}=\frac{(\check{\bm{v}}^{(r)} +\bm{\xi}(\check{\bm{v}}^{(r)}))_{m}} {\vert\check{\bm{v}}^{(r)}+\bm{\xi}(\check{\bm{v}}^{(r)})\vert_{m}}, \ \ \forall m.
\end{align}

We summarize the RNM-based algorithm for solving \eqref{Optprob_v} in Algorithm 1, where the stop criteria of RNM and FP are set as  $|f(\check{\bm{v}}^{(r+1)})-f(\check{\bm{v}}^{(r)})|<\delta_1$ and $|z^{(q+1)}-z^{(q)}|<\delta_2$, respectively, with the stop tolerances $\delta_1$ and $\delta_2$.
It is worth noting that introducing the diagonal loading matrix $\frac{\mu}{2}\bm{I}_{2M}$ in \eqref{NewtonEq2} does not destroy the superlinear convergence property of the pure Newton iteration when the desired local minimum is reached \cite{Absil2008Optimization}.
Therefore, the proposed RNM should have faster convergence speed than typical first-order RMO methods.

\subsection{Optimization of Receive Beamforming}
Fixing the variables $\bm{u}$, $\bm{v}$ and discarding irrelative constant items, we can express \eqref{Optprob1} with respect to $\bm{w}$ as
\begin{equation}
\label{Optprob_w}
\max_{\bm{w}}\frac{|\bm{w}^H \bm{\Phi}(\theta_0)\bm{u}|^2}{\bm{w}^H\bm{\Psi_u}\bm{w}+\sigma_n^2\bm{w}^H\bm{w}}
\end{equation}
Based on the principle of MVDR, we obtain the solution of the problem \eqref{Optprob_w} as
\begin{equation}
\label{eq:w_opt0}
\bm{w}=\alpha_w (\bm{\Psi_u}+\sigma_n^2\bm{I}_N)^{-1}\bm{\Phi}(\theta_0)\bm{u},
\end{equation}
where $\alpha_w \neq 0$ is a constant which does not affect the objective function value in \eqref{Optprob_w}.
For simplicity, we set $\alpha_w=1$. 
Hence the solution of \eqref{Optprob_w} is briefly expressed as
\begin{equation}
\label{eq:w_opt}
\bm{w}=(\bm{\Psi_u}+\sigma_n^2\bm{I}_N)^{-1}\bm{\Phi}(\theta_0)\bm{u}.
\end{equation}

\subsection{Optimization of Transmit Beamforming}
Fixing the variables $\bm{w}$, $\bm{v}$ and ignoring irrelative constant items, we can rewrite \eqref{Optprob1} with respect to $\bm{u}$ as
\begin{subequations}
	\label{Optprob_tTP}
	\begin{align}
	\label{Optprob_tTP1}
	\max_{\bm{u}} &\ \frac{|\bm{u}^H\bm{\Phi}^H(\theta_0)\bm{w}|^2} {\bm{u}^H(\bm{\Psi_w} +\sigma_n^2\bm{w}^H\bm{w})\bm{u}}   \\
	\label{Optprob_tTP2}
	\rm{s.t.} \ & \ \Vert\bm{u}\Vert_2^2=1
	\end{align}
\end{subequations}	
where $\bm{\Psi_w}=\sum_{i=1}^{I}\sigma_i^2 \bm{\Phi}^H(\theta_i)\bm{w}\bm{w}^H\bm{\Phi}(\theta_i)$ and the constraint $\Vert\bm{u}\Vert_2^2=1$ is invoked in the denominator of the cost function.
According to \eqref{eq:w_opt0}, we obtain the solution of \eqref{Optprob_tTP} without the constraint \eqref{Optprob_tTP2} as 
\begin{equation}
\label{eq:t_opt0}
\bm{u}=\alpha_u (\bm{\Psi_w}+\sigma_n^2\bm{w}^H\bm{w}\bm{I}_N)^{-1}\bm{\Phi}^H(\theta_0)\bm{w}
\end{equation}
where $\alpha_u \neq 0$ is a constant which does not affect the objective function value of \eqref{Optprob_tTP1}.
We can directly determine $\alpha_u$ from the constraint \eqref{Optprob_tTP2} and then yield the solution of \eqref{Optprob_tTP} as
\begin{equation}
\label{eq:t_optTP}
    \bm{u}=\frac{(\bm{\Psi_w}+\sigma_n^2\bm{w}^H\bm{w}\bm{I}_N)^{-1}\bm{\Phi}^H(\theta_0)\bm{w}} {\Vert(\bm{\Psi_w}+\sigma_n^2\bm{w}^H\bm{w}\bm{I}_N)^{-1}\bm{\Phi}^H(\theta_0)\bm{w}\Vert_2}.
\end{equation}

By solving the problems \eqref{Optprob_v}, \eqref{Optprob_w} and \eqref{Optprob_tTP} repeatedly until a termination criterion meets, we obtain the optimized $\bm{v}^{\star}$, $\bm{w}^{\star}$ and $\bm{u}^{\star}$. 
We summarize the proposed beamforming codesign algorithm in Algorithm 2, where $\delta_3$ is the stop tolerance and $P_{\max}$ is the maximum iteration number.

\begin{algorithm}[t!]
	\caption{Proposed Beamforming Codesign Algorithm.}
	\label{alg:Proposed}
	\begin{algorithmic}[1]
		\renewcommand{\algorithmicrequire}{\textbf{Input:}}
		\REQUIRE $\bm{u}^{0}$, $\bm{v}^{0}$, $\bm{w}^{0}$, $\bm{a}(\theta_i)$, $\bm{b}(\theta_i+\theta_{tr})$, $\sigma^2_i$, $\sigma^2_n$, $\bm{G}$. \\         
		\renewcommand{\algorithmicensure}{\textbf{Output:}}
		\ENSURE The beamformers $\bm{v}^{\star}$, $\bm{u}^{\star}$ and $\bm{w}^{\star}$.    
		\STATE Set $p=0$.                
		\WHILE{$p<P_{\max}$}
		\STATE Obtain $\bm{v}^{p+1}$ using Algorithm 1; 
		\STATE Obtain $\bm{w}^{p+1}$ using \eqref{eq:w_opt};
		\STATE Obtain $\bm{u}^{p+1}$ using \eqref{eq:t_optTP};
		\STATE If $|G(\bm{u}^{p+1},\bm{v}^{p+1},\bm{w}^{p+1})-G(\bm{u}^p,\bm{v}^p,\bm{w}^p)|<\delta_3$, go to Output; 
		Otherwise, $p=p+1$, go to Step 2;	
		\ENDWHILE	
	\end{algorithmic}
\end{algorithm}

\section{Convergence and complexity analysis}
\subsection{Convergence analysis}
In Algorithm 2, we alternately update $\bm{v}$, $\bm{w}$ and $\bm{u}$ through Algorithm 1, \eqref{eq:w_opt} and \eqref{eq:t_optTP}, respectively. 
Since the optimal $\bm{w}$ and $\bm{u}$ are explicitly given by \eqref{eq:w_opt} and \eqref{eq:t_optTP}, the update of $\bm{w}$ and $\bm{u}$ ensure the non-decrease of the objective function $G(\bm{u},\bm{v},\bm{w})$ in \eqref{Optprob1}. 
Note that $G(\bm{u},\bm{v},\bm{w})$ is necessarily upper bounded.
Therefore, if the update of $\bm{v}$ by Algorithm 1 still guarantee $G(\bm{u},\bm{v},\bm{w})$ is non-decreasing, Algorithm 2 necessarily converges to a finite value of $G(\bm{u},\bm{v},\bm{w})$ according to the monotone convergence theorem. 

Based on the aforementioned analysis, we need to prove that $G(\bm{u},\bm{v},\bm{w})$ is always non-decreasing after the update of Algorithm 1.
Before proceeding, we give the equivalent expression of the objective function $f(\bm{v})$ in \eqref{Dinkelbach_v3} and the projector $\mathcal{P}_{\bm{v}}(\bm{y})$ in real number form. 
The real number expression of $f(\bm{v})$ is
\begin{equation}
	\label{eq:opteq}
	\begin{split}
		\hat{f}(\hat{\bm{v}})=\hat{\bm{v}}^T\hat{\bm{H}}\hat{\bm{v}}+2\hat{\bm{v}}^T\hat{\bm{h}},
	\end{split}
\end{equation}
where $\hat{\bm{h}}=[\mathrm{Re}\{\tilde{\bm{h}}(z)\}^T,\mathrm{Im}\{\tilde{\bm{h}}(z)\}^T]^T$.

Note that the projector $\mathcal{P}_{\bm{v}}(\bm{y})$ can be described as
\begin{equation}
	\mathcal{P}_{\bm{v}}(\bm{y})=\bm{y}-\mathrm{Re}\{\bm{y}\odot \bm{v}^{*}\} \odot \bm{v}=\frac{1}{2}(\bm{y}-\mathrm{Diag}\{\bm{v}\odot\bm{v}\}\bm{y}^{*}).
\end{equation}
Then its equivalent real number expression is
\begin{equation}
	\label{eq:realPv}
	\begin{split}
		\hat{\mathcal{P}}_{\hat{\bm{v}}}(\hat{\bm{y}})=\frac{1}{2}(\hat{\bm{y}}- \hat{\bm{V}}\hat{\bm{y}})=\frac{1}{2}(\bm{I}_{2M}-\hat{\bm{V}})\hat{\bm{y}},
	\end{split}
\end{equation}
with $\hat{\bm{y}}=[\mathrm{Re}\{\bm{y}\}^T,\mathrm{Im}\{\bm{y}\}^T]^T$. 

According to \eqref{eq:realPv}, the relationship between Riemannian gradient and Euclidean gradient in \eqref{RiemGradf} can be expressed in the real number form as 
\begin{equation}
	\label{eq:hatgw}
	\hat{\bm{g}}=\hat{\mathcal{P}}_{\hat{\bm{v}}}(\hat{\bm{e}})=\frac{1}{2}(\bm{I}_{2M}- \hat{\bm{V}})\hat{\bm{e}},
\end{equation}
where $\hat{\bm{e}}=[\mathrm{Re}\{\triangledown f(\bm{v})\}^T, \mathrm{Im}\{\triangledown f(\bm{v})\}^T]^T$ is the real Euclidean gradient.
Based on the expression in \eqref{eq:hatgw}, we have derived that the real improved RND generated by \eqref{NewtowDirect} is always on the tangent space $\mathbb{T}_{\bm v}\mathcal{M}$ in Appendix \ref{sec:TM}, namely, 
\begin{equation}
	\label{eq:RNDonTS}
	\hat{\bm{\xi}}^{\star}=\hat{\mathcal{P}}_{\hat{\bm{v}}}(\hat{\bm{\xi}}^{\star})=\frac{1}{2}(\bm{I}_{2M}-\hat{\bm{V}})\hat{\bm{\xi}}^{\star}.
\end{equation}
Then the equation \eqref{NewtonEq2} can be equivalently expressed in the real number form as
\begin{equation}
	\label{eq:NewtonEquation2}
	\hat{\bm{g}}+\frac{1}{4}[(\bm{I}_{2M}-\hat{\bm{V}})(\hat{\bm{H}}-\hat{\bm{Q}})+\mu \bm{I}_{2M}](\bm{I}_{2M}-\hat{\bm{V}})\hat{\bm{\xi}} = 0.
\end{equation}

Inserting \eqref{eq:RNDonTS} into \eqref{eq:NewtonEquation2}, we have
\begin{equation}
	\hat{\bm{g}}+\frac{1}{4}[(\bm{I}_{2M}-\hat{\bm{V}})(\hat{\bm{H}}-\hat{\bm{Q}})(\bm{I}_{2M}-\hat{\bm{V}})+2\mu \bm{I}_{2M}]\hat{\bm{\xi}}=0.
\end{equation}
Therefore, we obtain another representation of the real improved RND by
\begin{equation}
	\label{eq:RND2}
	\hat{\bm{\xi}}^{\star}=-4\hat{\bm{H}}_H^{-1}\hat{\bm{g}} =-2\hat{\bm{H}}_H^{-1}(\bm{I}_{2M}-\hat{\bm{V}})\hat{\bm{e}},
\end{equation}
where $\hat{\bm{H}}_H=(\bm{I}_{2M}-\hat{\bm{V}})(\hat{\bm{H}}-\hat{\bm{Q}})(\bm{I}_{2M}-\hat{\bm{V}})+2\mu \bm{I}_{2M}$.
For ease of expression, denote $\hat{\bm{H}}_{H_0}=(\bm{I}_{2M}-\hat{\bm{V}})(\hat{\bm{H}}-\hat{\bm{Q}})(\bm{I}_{2M}-\hat{\bm{V}})$.
Then we have $\hat{\bm{H}}_H=\hat{\bm{H}}_{H_0}+2\mu \bm{I}_{2M}$.

Now, we first derive the convergence of the proposed RNM, and then deduce the convergence of Algorithm 1. 
Note that each iteration of RNM contains two steps: searching along the improved RND in \eqref{RiemaNewUpdate} and retraction in \eqref{RiemaProj}.
We provide the sufficient conditions to ensure that the objective function $f(\bm{v})$ is monotonically non-increasing in both steps, respectively.

\emph{Proposition IV.1}: Let $\lambda_{1}$ denote the maximum eigenvalue of $\hat{\bm{H}}-\frac{\hat{\bm{H}}_{H_0}}{2}$.
If $\mu>\lambda_{1}$, $f(\tilde{\bm{v}}^{(r+1)})\leq f(\check{\bm{v}}^{(r)})$ always holds.

\emph{Proof}: See Appendix \ref{sec:plemma1}. 

From Proposition IV.1, the improved RND ensures that $f(\bm{v})$ is always non-increasing with a properly selected $\mu$.
Subsequently, we dedicate to making $f(\bm{v})$ monotonically non-increasing in the retraction step.

For the convenience of description, we redefine $\lambda_v$ in \eqref{Dinkelbach_v3}
as $\lambda_v=\lambda_{p}+\lambda_c$ with $\lambda_{p}>0$ and $\lambda_c>0$.
Let $\lambda_{p} > \Vert\bm{h}(\theta_0)\Vert_2^2$ to ensure the positive definiteness of $\bar{\bm{H}}=\sum_{i=1}^{I}z\sigma_i^2\bm{H}_i-\bm{H}_0+\lambda_p\bm{I}_M$. 
Then we rephrase $\tilde{\bm{H}}=\sum_{i=1}^{I}z\sigma_i^2\bm{H}_i-\bm{H}_0+\lambda_v\bm{I}_M$ as $\tilde{\bm{H}}=\bar{\bm{H}}+\lambda_c\bm{I}_M$. 
With a proper selection of $\lambda_c$, the following proposition shows that $f(\bm{v})$ is always non-increasing in each retraction step.

\emph{Proposition IV.2}: Let $\lambda_{2}$ denote the largest eigenvalue of the matrix $\bar{\bm{H}}$. 
If $\lambda_{p}>\Vert\bm{h}(\theta_0)\Vert_2^2$ and $\lambda_c\geq \frac{M}{8}\lambda_2+\Vert\tilde{\bm{h}}(z)\Vert_2$ , $f(\check{\bm{v}}^{(r+1)})\leq f(\tilde{\bm{v}}^{(r+1)})$ always holds. 

\emph{Proof}: This proposition can be directly obtained according to Appendix B of \cite{Alhujaili2019Transmit}. 

Based on Proposition IV.1 and Proposition IV.2, we find that Algorithm 1 is convergent under the following conditions. 

\emph{Theorem IV.3}: Let $\{\bm{v}^{(q)}\}$ be a sequence generated by Algorithm 1 for solving \eqref{Optprob_v}. 
If $\mu>\lambda_{1}$, $\lambda_{p}>\Vert\bm{h}(\theta_0)\Vert_2^2$, and $\lambda_c\geq \frac{M}{8}\lambda_2+\Vert\tilde{\bm{h}}(z)\Vert_2$, the sequence of the Dinkelbach variable $\{z^{(q)}\}$ is monotonically non-decreasing and converges to a finite value and then Algorithm 1 is convergent. 

\emph{Proof}: See Appendix \ref{sec:plemma3}. 

According to Theorem IV.3, we know that the updated $\bm{v}^{p+1}$ yielded by Algorithm 1 ensures the monotonically non-decreasing of $G(\bm{u}^{p},\bm{v},\bm{w}^{p})$ in \eqref{Optprob1}. 
Therefore, we ensure that $G(\bm{u},\bm{v},\bm{w})$ is non-decreasing at each iteration step of Algorithm 2. 
Since $G(\bm{u},\bm{v},\bm{w})$ is necessarily upper bounded, Algorithm 2 converges to a finite value $G^\star$ according to the monotone convergence theorem.

\vspace{-2mm}
\subsection{Computational complexity analysis}
The computational complexity of Algorithm 2 depends on the obtaining of $\bm{v}$, $\bm{w}$ and $\bm{u}$. 
For Algorithm 1 to update $\bm{v}$, the computational cost is dominated by the iteratively solving of the subproblem \eqref{Dinkelbach_v3}.
Denote $N_{\text{Dinv}}$ and $N_{\text{RNM}}$ as the maximum iteration number of the Dinkelbach transform and the RNM, respectively.
At each iteration of RNM, the cost depends on the calculation of RND in \eqref{NewtowDirect}, which equals $\mathcal{O}(M^3)$. 
Hence the total cost of Algorithm 1 is about $\mathcal{O}(N_{\text{Dinv}}N_{\text{RNM}}M^3)$.
For the update of $\bm{w}$ in \eqref{eq:w_opt}, the computational cost is mainly contributed by the generation and the inversion of $\bm{\Psi_u}$, which equal $\mathcal{O}(IMN+IN^2)$ and $\mathcal{O}(N^3)$, respectively. 
Thus the total cost is about $\mathcal{O}(IMN+IN^2+N^3)$. 
As for the update of $\bm{u}$ in \eqref{eq:t_optTP}, the cost is similar to that in \eqref{eq:w_opt} and is about $\mathcal{O}(IMN+IN^2+N^3)$.
In practical RIS-assisted array radars, we usually have $I,N < M$.
Therefore, the total computational cost of Algorithm 2 is dominated by the update of $\bm{v}$ and is about $\mathcal{O}(N_{\text{Dinv}}N_{\text{RNM}}M^3)$.

\section{Numerical Results}
In this section, we present several numerical experiments to validate the convergence and superior performance of the proposed codesign algorithm and to quantitatively analyze the impact of system parameters.
We set the transceiver equipped with isotropic antennas and half-wavelength element spacing.
The target SNR and the interference-to-noise ratio (INR) of the $i$-th interference are defined as $\mathrm{SNR}=10\log_{10}\alpha_0^2/\sigma_n^2$ and $\mathrm{INR}_i=10\log_{10}\alpha_i^2/\sigma_n^2$, respectively.
The distance-dependent path loss model is given by
\begin{equation}
	\label{eq:largescaleloss}
	C=C_0(\frac{D}{D_0})^{-\eta},
\end{equation}
where $C_0$ is the path loss at the reference distance $D_0$, $D$ is the distance between the array and RIS, and $\eta$ is the path loss exponent.
The Rician fading channel model is adopted to account for small-scale fading\cite{Wu2019Intelligent}, which is given by
\begin{equation}
	\label{eq:RicianChannel}
	\bm{G}=\sqrt{\frac{K_\text{R}}{1+K_\text{R}}}\bm{G}^{\text{LoS}}+\sqrt{\frac{1}{1+K_\text{R}}}\bm{G}^{\text{NLoS}}
\end{equation}
where $K_\text{R}$ is the Rician factor, $\bm{G}^{\text{LoS}}$ represents the deterministic LoS component and $\bm{G}^{\text{NLoS}}$ represents the Rayleigh fading component whose entries are independent and identically distributed circularly symmetric complex Gaussian random variable with zero mean and unit variance.
Then $\bm{G}$ is multiplied by $\sqrt{C}$.
The potential target is located at the look direction $30^\circ$ with $\mathrm{SNR}=10\mathrm{dB}$ and all interferences have identical power.
Among all interferences, we assume that three of them are close to each other and thus form a interference cluster within the look direction region of $2^\circ$.
Unless otherwise stated, we set $N=5$, $\theta_{tr}=20^\circ$,  $\mathrm{INR}=30\mathrm{dB}$, $D_0=1$m, $D=3$m, $C_0=-30\mathrm{dB}$, $\eta=2.2$ and $K_\text{R}=0.5$.

\vspace{-2mm}
\subsection{Convergence Analysis}

\begin{figure}[t!]
	\vspace{-4mm}
	\centering
	\includegraphics[width=0.5\linewidth]{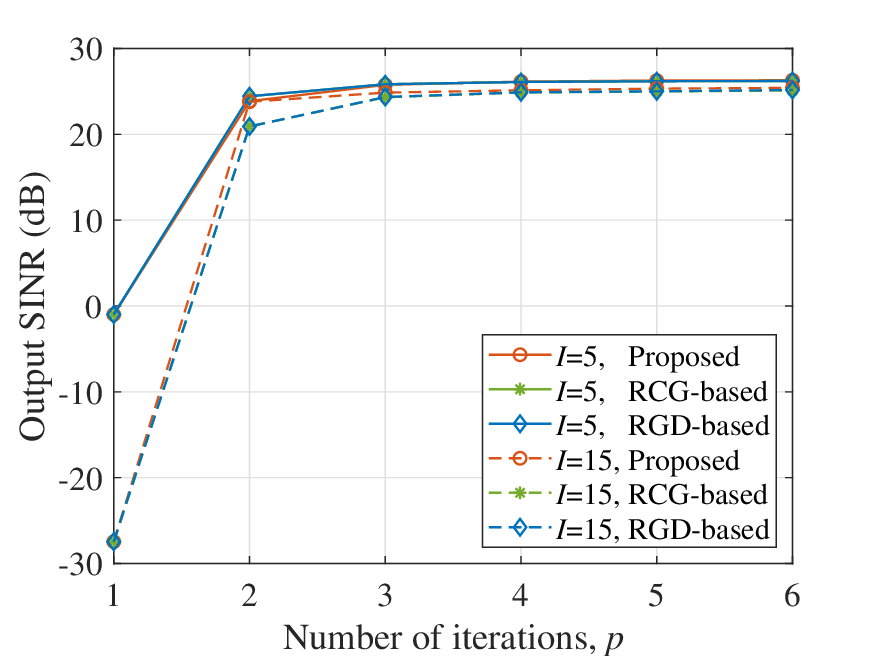}
	\caption{Output SINR v.s. the number of iterations. } 
	\label{fig:conver}
	\vspace{-4mm}
\end{figure}

\begin{figure}[t!]
	\centering
	\subfloat[]{\includegraphics[width=0.5\linewidth]{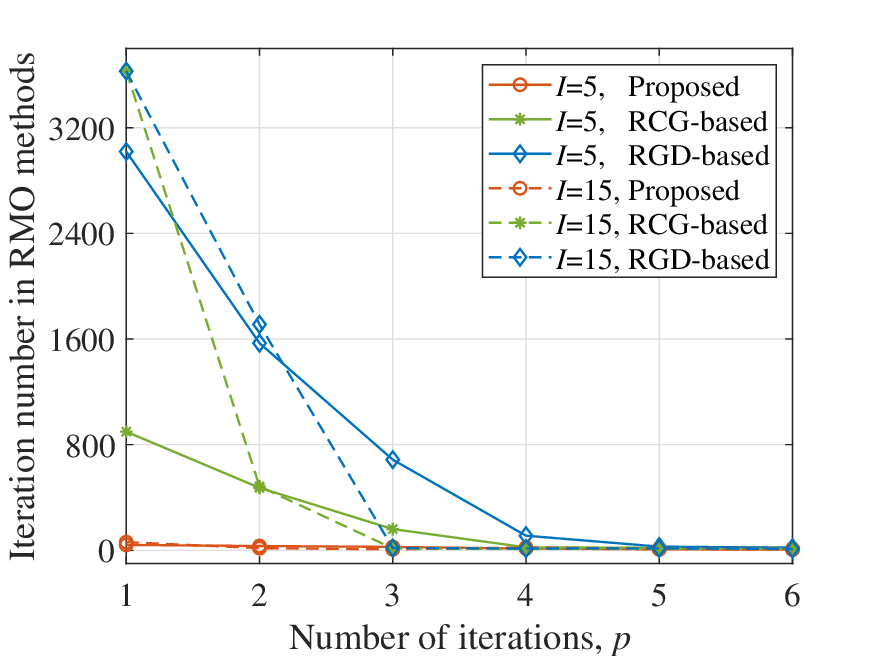}\label{fig:convtime}}  
	\subfloat[]{\includegraphics[width=0.5\linewidth]{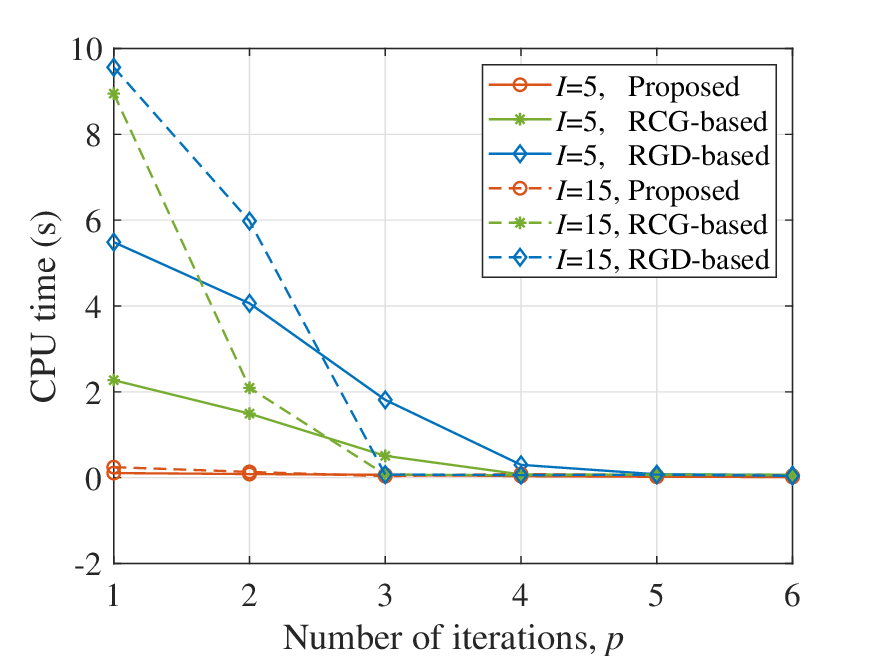}\label{fig:convnum}}
	\caption{The performance of three RMO methods in each update step of $\bm{v}$. (a) iteration number, (b) CPU time consumption.}
	\label{fig:conver1}
	\vspace{-7mm}
\end{figure}

\begin{figure}[t!]
	\centering
	\subfloat[]{\includegraphics[width=0.5\linewidth]{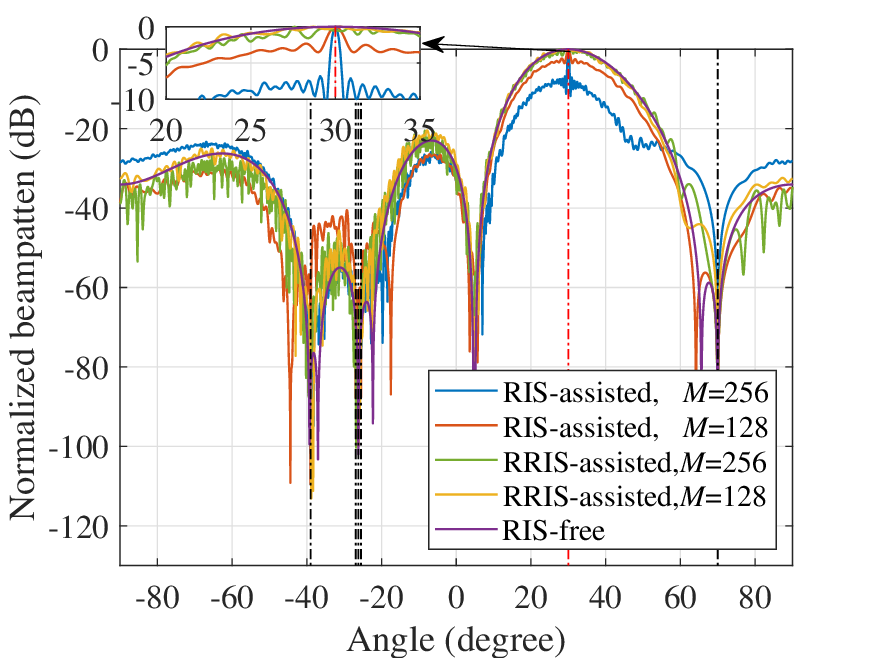}\label{fig:bpinf5}}  
	\subfloat[]{\includegraphics[width=0.5\linewidth]{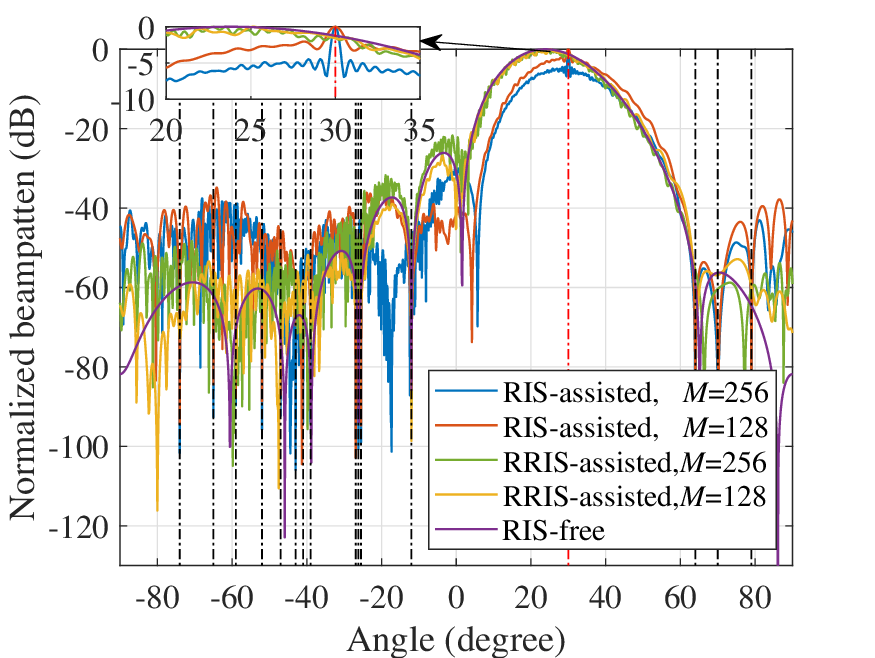}\label{fig:bpinf15}}
	\caption{Joint transmit and receive beampatterns for different kinds of array radars. (a) $I=5$, (b) $I=15$.}
	\label{fig:BF}
	\vspace{-7mm}
\end{figure}

To examine the convergence of Algorithm 2, we replace the RNM for solving \eqref{Dinkelbach_v3} with the RCG and RGD \cite{Li2020Riemannian}, respectively, and thus yield the RCG- and RGD-based algorithms as benchmarks.
We compare the convergence speed and CPU time consumption of all three algorithms by using a standard PC with AMD Ryzen 7 5800X CPU and 32 GB RAM.
Fig. \ref{fig:conver} presents the output SINR against the number of iterations for Algorithm 2 using different RMO solvers, where $M=128$, $I=5$ or $I=15$.
All algorithms use the same initialization and the maximum iteration number equals $P_{\rm max}=9$.
We find that all three algorithms yield non-decreasing output SINR as the number of iterations increases and almost converge to a fixed value after $4$ iterations.

Fig. \ref{fig:conver1} displays the performance of three RMO methods in the update step of $\bm{v}$.
As seen in Fig. \ref{fig:conver1}(a), both RCG and RGD consume much greater iteration numbers than the RNM in the first three updates of $\bm{v}$, and then dramatically decrease after $p \geq 4$.
Concretely, at the first update of $\bm{v}$, the iteration numbers are respectively about 40, 900 and 3000 in RNM, RCG and RGD when $I=5$, and increase to 60, 3600 and 3600 when $I=15$.
The proposed RNM converges significantly faster than RCG and RGD due to the quadratic convergence rate.
From Fig. \ref{fig:conver1}(b), we observe that the RNM consumes much less CPU time than RCG and RGD.
At the first update, the CPU times consumed by RNM, RCG and RGD are respectively about 0.1s, 2.3s and 5.5s when $I=5$, and increase to 0.2s, 8.9s and 9.6s when $I=15$.
Even though the RNM theoretically has higher computational complexity than RCG and RGD at each iteration, its faster convergence speed and avoidance of line search effectively remedy this shortcoming in the updates of $\bm{v}$.
Actually, the total CPU times of the proposed, RCG-based and RGD-based algorithms equal 0.3s, 4.4s and 11.7s when $I=5$, and 0.5s, 11.2s and 15.7s when $I=15$, respectively.
Therefore, the proposed algorithm is much computationally efficient to jointly design the transmit beamforming and RIS-assisted receive beamforming.

\vspace{-3mm}
\subsection{Beamforming Performance}
This subsection tests the performance improvement offered by the RIS.
To show the effectiveness of deploying the RIS and the proposed codesign algorithm, we also present the performance of conventional array radar without RIS (abbreviated as RIS-free) and RIS-assisted one with random RIS reflection coefficients (abbreviated as RRIS-assisted) as benchmarks.

We first display the joint transmit and receive beampatterns of RIS-assisted, RRIS-assisted and RIS-free array radars in Fig. \ref{fig:BF}.
In Fig. \ref{fig:BF}(a), there exist a cluster of interferences and 2 individual point interferences.
The cluster consists of 3 point interferences located at the directions $-25.5^\circ$, $-26.2^\circ$, and $-26.9^\circ$, and two point interferences are located at the directions $-39^\circ$ and $70^\circ$.
Compared with RRIS-assisted and RIS-free radars, the RIS-assisted ones yield a narrower mainlobe at the target direction and lower sidelobes, which should result in a better SINR performance.
In fact, the output SINRs of RIS-assisted radars equal $29.9\mathrm{dB}$ and $26.7\mathrm{dB}$ with 256- and 128-element RISs, respectively, while those of RRIS-assisted and RIS-free radars are only about $23.7\mathrm{dB}$.
More RIS elements lead to higher output SINRs in the RIS-assisted radars, but random RISs do not offer the SINR improvement. 

In Fig. \ref{fig:BF}(b), another $10$ interferences are added with the look directions being $79^\circ$, $64^\circ$, $-12^\circ$, $-41^\circ$, $-43^\circ$, $-47^\circ$, $-52^\circ$, $-59^\circ$, $-65^\circ$ and $-74^\circ$.
We observe that only the RIS-assisted radars form correct nulls at all interference directions and thus will have the best SINR performance.
Actually, the output SINRs of 256- and 128-element RIS-assisted radars equal $28.3\mathrm{dB}$ and $25.7\mathrm{dB}$,  respectively, while those of 256- and 128-element RRIS-assisted radars and RIS-free radar are about $20.3\mathrm{dB}$, $20.8\mathrm{dB}$ and $21.0\mathrm{dB}$.
Compared to the RIS-assisted radars, the RRIS-assisted and RIS-free radars suffer a more severe SINR reduction.
Moreover, random RISs produce a negative impact on the output SINR in this case.
Therefore, deploying the RIS optimized by the proposed codesign algorithm remarkably improves the SINR performance, especially when a great number of interferences appear.

\begin{figure}[t!]
	\vspace{-3mm}
	\centering
	\subfloat[]{\includegraphics[width=0.5\linewidth]{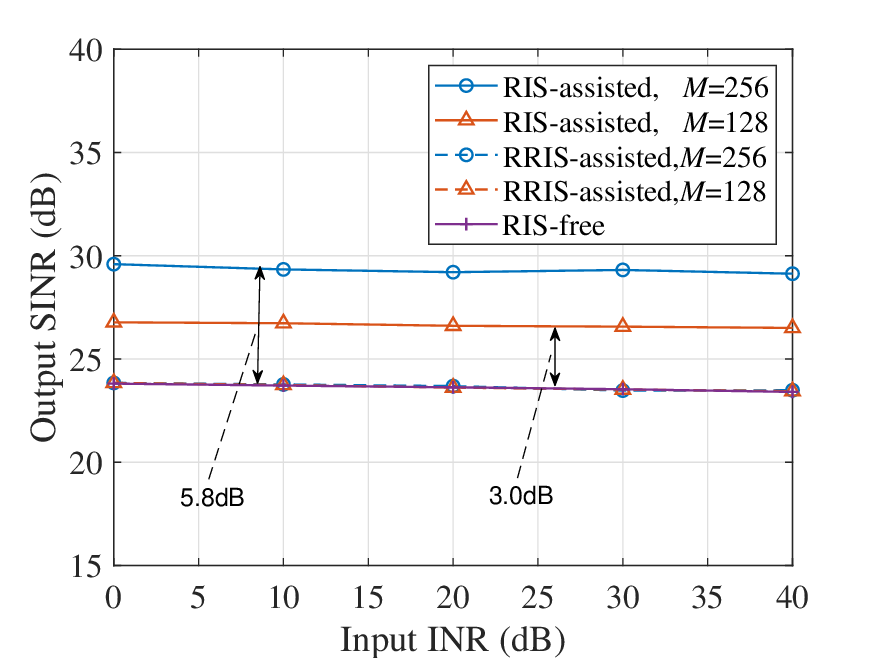}\label{fig:inrsinr5}}  
	\subfloat[]{\includegraphics[width=0.5\linewidth]{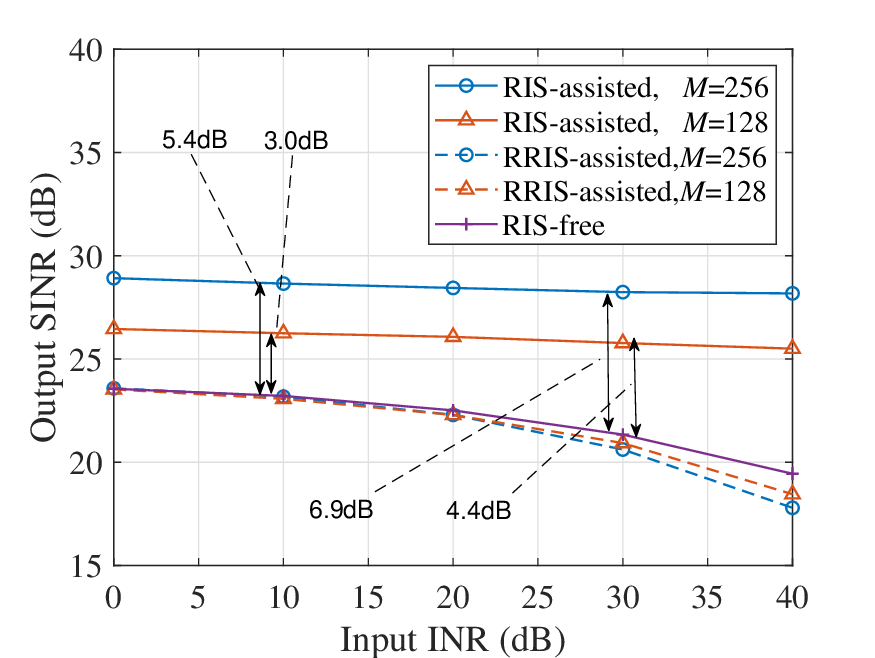}\label{fig:inrsinr15}}
	\caption{Output SINR versus input INR. (a) $I=5$, (b) $I=15$. }
	\label{fig:inrsinr}
	\vspace{-7mm}
\end{figure}

Next, we exhibit the superior SINR performance of RIS-assisted array radar under different scenarios.
Unless otherwise stated, the output SINR is obtained by averaging 50 independent runs.
Fig. \ref{fig:inrsinr} shows the output SINR versus the input INR with different RIS element numbers.
From Fig. \ref{fig:inrsinr}\subref{fig:inrsinr5} with 5 interferences existing, we find that the SINRs of all five radars are almost not affected by the input INR.
Compared to the RIS-free radar, the output SINRs are improved by 5.8dB and 3.0dB with the aid of 256- and 128-element RISs, respectively, but the RRIS-assisted radars do not offer any SINR improvement.
In contrary, when the interference number increases to 15, the output SINR of RIS-assisted radars are nearly unchanged with the increase of input INR, but those of RRIS-assisted and RIS-free ones gradually decrease, as seen in Fig. \ref{fig:inrsinr}\subref{fig:inrsinr15}.
Then the SINR improvement of RIS-assisted radars becomes larger when the input INR grows up.
On the other side, the RRIS-assisted radars have inferior performance to the RIS-free one, and their SINRs decrease more rapidly for a larger-size RIS.
Hence deploying a random RIS cannot enhance the capability of suppressing strong interferences, but has a destructive effect.

\begin{figure}[t!]
	\vspace{-4mm}
	\centering
	\includegraphics[width=0.5\linewidth]{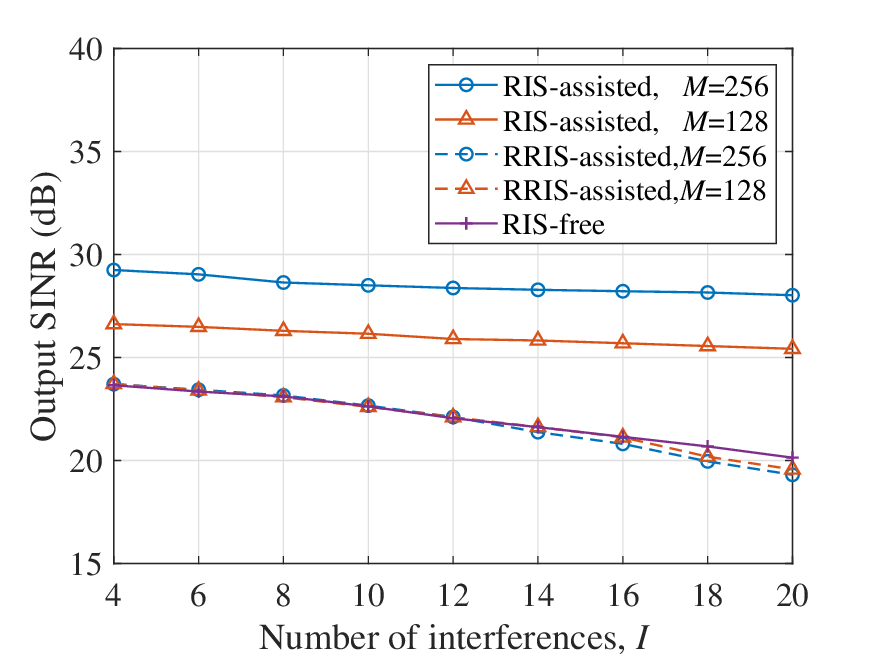}
	\caption{Output SINR versus the number of interferences $I$.}
	\label{fig:Nisinr}
	\vspace{-6mm}
\end{figure}

We further examine the relationship between the output SINR and the number of interferences.
Fig. \ref{fig:Nisinr} discloses that the output SINR of RIS-assisted radars are nearly unaffected by the number of interferences.
However, the RRIS-assisted and RIS-free radars suffer an evident performance degradation as the number of interferences increases, especially when the interference number  is greater than the DoFs of active array ($2N-1=9$).
The SINR decrease of RRIS-assisted radars is larger than that of RIS-free radar, and the RRIS-assisted radar with 256 elements suffers larger SINR decrease than that with 128 elements.
This is because the deployed RIS greatly increases system DoFs and thus offers the capability of forming more nulls to suppress a large number of interferences.
However, deploying a random RIS not only fails to increase system DoFs, but also 
disrupts the original ability of interference suppression. 
Only equipping an optimized RIS can improve the SINR performance and enable conventional array radars to suppress more and stronger interferences.

\begin{figure}[t!]
	\centering
	\includegraphics[width=0.5\linewidth]{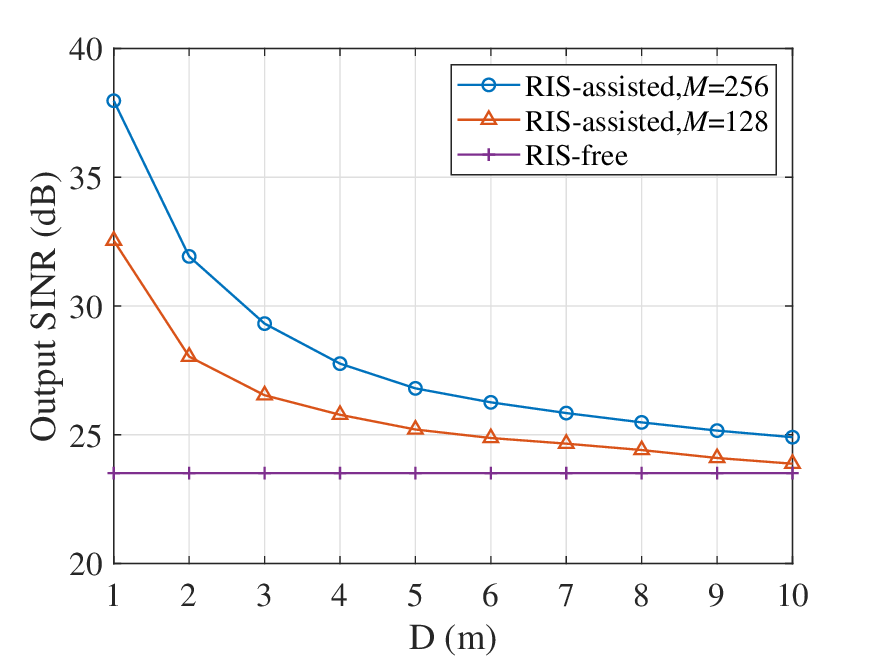}
	\caption{Output SINR versus the distance between radar and RIS.}
	\label{fig:Dsinr}
	\vspace{-7mm}
\end{figure}

\subsection{Effect of System Parameters}
This subsection explores the effect of system parameters on the output SINR, including the distance between radar and RIS, the number of RIS elements, and the number of radar array elements.
Fig. \ref{fig:Dsinr} shows the influence of the distance between radar and RIS on the output SINR. 
We observe that shorter distance always leads to larger SINR. 
When the distance is $D=1\mathrm{m}$, the output SINRs are $38.0\mathrm{dB}$, $32.5\mathrm{dB}$ and $28.4\mathrm{dB}$ with RIS containing 256, 128, and 64 elements, respectively. 
However, when the distance increases to $D=3\mathrm{m}$, the output SINRs drop to $29.3\mathrm{dB}$, $26.5\mathrm{dB}$ and $25.1\mathrm{dB}$, respectively.
This phenomenon results from the multiplication fading effect on the NLoS receive path loss and thus weakens the interference suppression capability provided by the RIS.
This path loss can be reduced by deploying the RIS close to the radar transceiver or the target \cite{Buzzi2022Foundations}.
Therefore, we should deploy the RIS as close as possible to the radar transceiver for better SINR performance. 
In fact, there exist two approaches to remedy this shortcoming.
The one is to increase the number of RIS elements, as is revealed in Fig. \ref{fig:Dsinr}.
Thanks to the low-cost property, we can employ the RIS with thousands of elements.
Then the path loss is partially compensated by the passive beamforming gain.
The other is to replace the passive RIS with an active RIS.
Since the active RIS has the capability of amplifying reflecting signals \cite{Zhang2023ActiveRIS}, it is naturally applied to remedy severe path loss. 
In fact, we have preliminarily verified that the active RIS can effectively improve the SINR performance in \cite{Feng2023Joint}.

\begin{figure}[t!]
	\centering
	\includegraphics[width=0.5\linewidth]{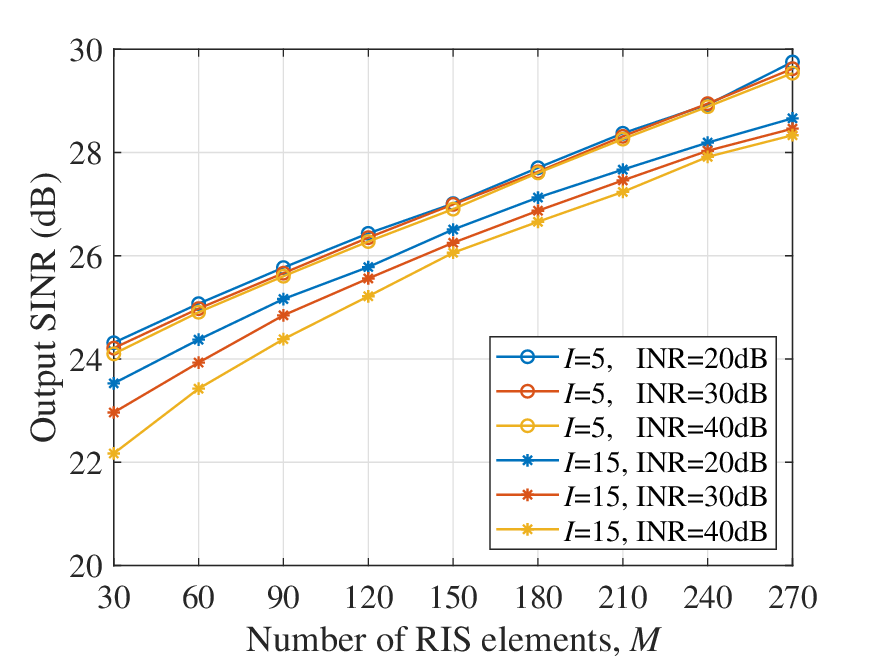}
	\caption{Output SINR versus the number of RIS elements.}
	\label{fig:Nrissinr}
	\vspace{-4mm}
\end{figure}

\begin{figure}[t!]
	\centering
	\includegraphics[width=0.5\linewidth]{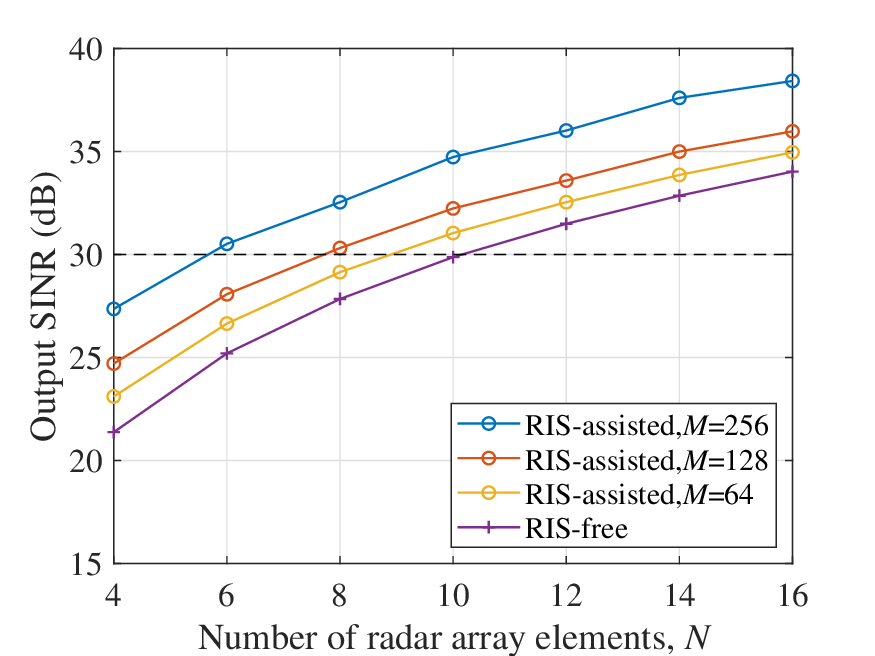}
	\caption{Output SINR v.s. the number of radar array elements.}
	\label{fig:Nasinr}
	\vspace{-6mm}
\end{figure}

Fig. \ref{fig:Nrissinr} illustrates the output SINR versus the number of RIS elements under different input INRs.
With the interference numbers $I=5$ and $I=15$, the output SINR always becomes higher with the increase of RIS element number and the SINR improvement is basically proportional to the number of RIS elements.
When $I=5$, the input INR has slight influence on the output SINR.
In contrast, when the interference number increases to $I=15$, a larger input INR always results in a lower output SINR, but this SINR performance difference is gradually alleviated with the increase of RIS element number.
That is because the DoFs of small-scale RIS-assisted radar are not sufficient enough to suppress strong interferences when plentiful interferences exist.
Hence we can deploy a large-scale RIS to strengthen the ability of strong interference suppression.

Fig. \ref{fig:Nasinr} presents the output SINR versus the number of radar array elements with different RIS element numbers, where $I=5$.
Not surprisingly, the output SINR becomes higher as the number of radar array elements increases in all four radars.
On the other hand, with a fixed radar array, more RIS elements always result in better SINR performance. 
Consequently, there exists a compromise between the number of RIS elements and the number of radar array elements in RIS-assisted array radar design.
As shown in Fig. \ref{fig:Nasinr}, to ensure the output SINR achieving $30\mathrm{dB}$, the RIS-free radar requires the active array containing at least 11 elements, while the RIS-assisted radar requires only 9, 8 and 6 array elements when the RIS element numbers equal 64, 128 and 256, respectively.
With the number of RIS elements increasing, the required number of radar array elements gradually decreases.
In practice, each radar array element needs to be equipped with a radio frequency and data acquisition channel, and thus fewer elements means lower cost and hardware complexity.
Therefore, the assistance of RIS provides a promising solution to enhance the flexibility of array radar design and to reduce the cost and complexity.

\section{Conclusion}
This paper provided a RIS-assisted receive mode to promote the ability of interference suppression for active array radars.
Under the total power constraint of transmit beamformer and the unimodular constraints of RIS reflection coefficients, we proposed a codesign algorithm of the transmit beamforming and RIS-assisted receive beamforming to maximize the output SINR.
We tackled the beamforming codesign problem under the alternating minimization framework, where the RIS reflection coefficients are optimized by exploiting the Dinkelbach transform and the proposed second-order RNM, and both the transmit and receive beamformers have closed-form optimal solutions. 
Moreover, we derived the convergence of the proposed codesign algorithm by deducing the explicit convergence condition of RNM.
Benefiting from the quadratic convergent rate and the avoidance of line search, the proposed codesign algorithm is significantly more computationally efficient than those using first-order RMO solvers.

Extensive experiment results demonstrated that deploying a closely-spaced RIS provides the capability of remarkably improving the interference suppression performance, especially for a great number of strong interferences appearing, and the output SINR improvement is proportional to the number of RIS elements.
Moreover, the assistance of RIS offers a promising approach to enhance the flexibility of system design and to reduce the cost and hardware complexity since there exists a compromise between the number of RIS elements and the number of radar array elements in RIS-assisted array radars.
It is worth pointing out that the RIS-induced NLoS path always suffers from the multiplication fading effect and the path loss dramatically grows up when the RIS moves away from the radar array.
To remedy this shortcoming, we consider to deploy several RISs at different locations in future works.

\appendix
\section{Appendix}
\subsection{Proof of the Improved RND in \eqref{NewtowDirect} Always on the Tangent Space. \label{sec:TM}}
\emph{Proof}: Projecting $\hat{\bm{\xi}}^{\star}$ onto the tangent space $\mathbb{T}_{\bm v}\mathcal{M}$, we have
\begin{equation}
	\label{eq:IRNDTS}
	\begin{split}
		&\frac{1}{2}(\bm{I}_{2M}-\hat{\bm{V}})\hat{\bm{\xi}}^{\star} \\
		=&-\frac{1}{2}(\bm{I}_{2M}-\hat{\bm{V}})[(\bm{I}_{2M}-\hat{\bm{V}})(\hat{\bm{H}}-\hat{\bm{Q}})+\mu \bm{I}_{2M}]^{-1}(\bm{I}_{2M	}-\hat{\bm{V}})\hat{\bm{e}} \\
		=&-\frac{1}{2}(\bm{I}_{2M}-\hat{\bm{V}}) \left\{ \frac{1}{\mu}\bm{I}_{2M}-\frac{1}{\mu}(\bm{I}_{2M}-\hat{\bm{V}})(\hat{\bm{H}}-\hat{\bm{Q}})\right.\\
		&\left.[\mu\bm{I}_{2M}+(\bm{I}_{2M}-\hat{\bm{V}})(\hat{\bm{H}}-\hat{\bm{Q}})]^{-1} \right\} (\bm{I}_{2M}-\hat{\bm{V}})\hat{\bm{e}} \\
		=&-\left\{\frac{1}{\mu}\bm{I}_{2M}-\frac{1}{\mu}(\bm{I}_{2M}-\hat{\bm{V}})(\hat{\bm{H}}-\hat{\bm{Q}})\right. \\
		&\left. [\mu\bm{I}_{2M}+(\bm{I}_{2M}-\hat{\bm{V}})(\hat{\bm{H}}-\hat{\bm{Q}})]^{-1} \right\} (\bm{I}_{2M}-\hat{\bm{V}}) \hat{\bm{e}} \\
		=&-[(\bm{I}_{2M}-\hat{\bm{V}})(\hat{\bm{H}}-\hat{\bm{Q}})+\mu \bm{I}_{2M}]^{-1}(\bm{I}_{2M}-\hat{\bm{V}})\hat{\bm{e}}=\ \hat{\bm{\xi}}^{\star}.
	\end{split}
\end{equation}
where the first equation is obtained by invoking \eqref{HessR}, \eqref{NewtowDirect} and \eqref{eq:hatgw}, the second equation is obtained by employing the matrix inversion lemma, the third equation is yielded by using the orthogonal projection property of $\mathcal{P}_{\hat{\bm{v}}}(\cdot)$, and the fourth equation is yielded by using the matrix inversion lemma again.

From \eqref{eq:IRNDTS}, we find that projecting $\hat{\bm{\xi}}^{\star}$ onto the tangent space $\mathbb{T}_{\bm v}\mathcal{M}$ equals itself. 
Therefore, $\hat{\bm{\xi}}^{\star}$ is on the tangent space.

\vspace{-10mm}
\subsection{Proof of Proposition IV.1 \label{sec:plemma1}}
\emph{Proof}: The real number form of \eqref{RiemaNewUpdate} is
\begin{equation}
	\bar{\bm{v}}^{(r+1)}=\hat{\bm{v}}^{(r)}+\hat{\bm{\xi}}^{\star}
\end{equation}
where $\bar{\bm{v}}^{(r+1)}=[\mathrm{Re}\{\tilde{\bm{v}}^{(r+1)}\}^T, \mathrm{Im}\{\tilde{\bm{v}}^{(r+1)}\}^T]^T$ and $\hat{\bm{v}}^{(r)}=[\mathrm{Re}\{\check{\bm{v}}^{(r)}\}^T, \mathrm{Im}\{\check{\bm{v}}^{(r)}\}^T]^T$. 
Inserting $\bar{\bm{v}}^{(r+1)}$ into $\hat{f}(\bm{v})$ yields
\begin{equation}
	\begin{split}
		\hat{f}(\bar{\bm{v}}^{(r+1)})&=(\hat{\bm{v}}^{(r)}+\hat{\bm{\xi}^{\star}})^T \hat{\bm{H}}(\hat{\bm{v}}^{(r)}+\hat{\bm{\xi}^{\star}})+2(\hat{\bm{v}}^{(r)}+\hat{\bm{\xi}^{\star}})^T\hat{\bm{h}}\\
		&=\hat{f}(\hat{\bm{v}}^{(r)})+2(\hat{\bm{\xi}}^{\star})^T(\hat{\bm{H}}\hat{\bm{v}}^{(r)}+\hat{\bm{h}})+(\hat{\bm{\xi}}^{\star})^T\hat{\bm{H}}\hat{\bm{\xi}^{\star}}.
	\end{split}
\end{equation}
Then, we have
\begin{equation}
	\label{eq:deltaf1}
	\begin{split}
		\hat{f}(\hat{\bm{v}}^{(r)})-\hat{f}(\bar{\bm{v}}^{(r+1)})=&-2(\hat{\bm{\xi}}^{\star})^T(\hat{\bm{H}}\hat{\bm{v}}^{(r)}+\hat{\bm{h}}) -(\hat{\bm{\xi}}^{\star})^T\hat{\bm{H}}\hat{\bm{\xi}^{\star}} \\
		=&-2(\hat{\bm{\xi}}^{\star})^T\hat{\bm{e}} -(\hat{\bm{\xi}}^{\star})^T\hat{\bm{H}}\hat{\bm{\xi}^{\star}}.
	\end{split}
\end{equation}
Inserting \eqref{eq:RND2} into \eqref{eq:deltaf1}, we obtain
\begin{equation}
	\label{eq:deltaf2}
	\begin{split}
		\hat{f}(\hat{\bm{v}}^{(r)})-&\hat{f}(\bar{\bm{v}}^{(r+1)})=
		4\hat{\bm{e}}^T(\bm{I}_{2M}-\hat{\bm{V}})\hat{\bm{H}}_H^{-1}\hat{\bm{e}} \\
		&\quad -4\hat{\bm{e}}^T(\bm{I}_{2M}-\hat{\bm{V}})\hat{\bm{H}}_H^{-1}\hat{\bm{H}}\hat{\bm{H}}_H^{-1}(\bm{I}_{2M}-\hat{\bm{V}})\hat{\bm{e}} .
	\end{split}
\end{equation}

Since $\hat{\bm{H}}_H=\hat{\bm{H}}_{H_0}+2\mu \bm{I}_{2M}$, we have 
\begin{equation}
	\label{eq:PD1}
	\begin{split}
		&\hat{\bm{e}}^T(\bm{I}_{2M}-\hat{\bm{V}})\hat{\bm{H}}_H^{-1}\hat{\bm{e}}\\
		&=\hat{\bm{e}}^T(\bm{I}_{2M}-\hat{\bm{V}})[\hat{\bm{H}}_{H_0}+2\mu \bm{I}_{2M}]^{-1}\hat{\bm{e}} \\
		&=\hat{\bm{e}}^T(\bm{I}_{2M}-\hat{\bm{V}})[\frac{1}{2\mu}\bm{I}_{2M}-\frac{1}{4\mu^2}(\bm{I}_{2M}+\frac{1}{2\mu}\hat{\bm{H}}_{H_0})^{-1}\hat{\bm{H}}_{H_0}]\hat{\bm{e}} \\
		&=\hat{\bm{e}}^T(\bm{I}_{2M}-\hat{\bm{V}})[\frac{1}{4\mu}\bm{I}_{2M}-\frac{1}{8\mu^2}(\bm{I}_{2M}+\frac{1}{2\mu}\hat{\bm{H}}_{H_0})^{-1}\hat{\bm{H}}_{H_0}]	\\
		& \ \ \ \ (\bm{I}_{2M}-\hat{\bm{V}})\hat{\bm{e}} \\
		&=\frac{1}{2}\hat{\bm{e}}^T(\bm{I}_{2M}-\hat{\bm{V}})[\hat{\bm{H}}_{H_0}+2\mu \bm{I}_{2M}]^{-1}(\bm{I}_{2M}-\hat{\bm{V}})\hat{\bm{e}} \\
		&=\frac{1}{2}\hat{\bm{e}}^T(\bm{I}_{2M}-\hat{\bm{V}})\hat{\bm{H}}_H^{-1}(\bm{I}_{2M}-\hat{\bm{V}})\hat{\bm{e}} .\\
	\end{split}
\end{equation}
where the second and the fourth equations are both obtained by employing the matrix inversion lemma, and the third equation is obtained according to $\hat{\bm{H}}_{H_0}\hat{\bm{e}}=\frac{1}{2}\hat{\bm{H}}_{H_0}(\bm{I}_{2M}-\hat{\bm{V}})\hat{\bm{e}}$ and $(\bm{I}_{2M}-\hat{\bm{V}})\hat{\bm{e}}= \frac{1}{2}(\bm{I}_{2M}-\hat{\bm{V}})(\bm{I}_{2M}-\hat{\bm{V}})\hat{\bm{e}}$.

Inserting \eqref{eq:PD1} into \eqref{eq:deltaf2}, we have
\begin{equation}
	\label{eq:deltaf3}
	\begin{split}
		\hat{f}&(\hat{\bm{v}}^{(r)})-\hat{f}(\bar{\bm{v}}^{(r+1)})\\
		&=2\hat{\bm{e}}^T(\bm{I}_{2M}-\hat{\bm{V}})\hat{\bm{H}}_H^{-1}(\bm{I}_{2M}-\hat{\bm{V}})\hat{\bm{e}} \\
		& \ \ \ \ -4\hat{\bm{e}}^T(\bm{I}_{2M}-\hat{\bm{V}})\hat{\bm{H}}_H^{-1}\hat{\bm{H}}\hat{\bm{H}}_H^{-1}(\bm{I}_{2M}-\hat{\bm{V}})\hat{\bm{e}} \\
		&=2\bm{e}_H^T(\hat{\bm{H}}_H -2\hat{\bm{H}})\bm{e}_H \\
		&=4\bm{e}_H^T(\mu\bm{I}_{2M}-\hat{\bm{H}}+\frac{\hat{\bm{H}}_{H_0}}{2})\bm{e}_H,
	\end{split}
\end{equation}
where $\bm{e}_H=\bm{H}_H^{-1}(\bm{I}_{2M}-\hat{\bm{V}})\hat{\bm{e}}$. 
Note that $\mu\bm{I}_{2M}-(\hat{\bm{H}}-\frac{\hat{\bm{H}}_{H_0}}{2})$ is positive definite if $\mu>\lambda_{1}$, in which $\lambda_{1}$ is the maximum eigenvalue of $\hat{\bm{H}}-\frac{\hat{\bm{H}}_{H_0}}{2}$. 
In this case, $\hat{f}(\hat{\bm{v}}^{(r)})-\hat{f}(\bar{\bm{v}}^{(r+1)}) \geq 0$, namely, $f(\tilde{\bm{v}}^{(r+1)}) \leq f(\check{\bm{v}}^{(r)})$.

\subsection{Proof of Theorem IV.3 \label{sec:plemma3}}
\emph{Proof}: From Proposition IV.1 and Proposition IV.2, we have $f(\check{\bm{v}}^{(r+1)})\leq f(\check{\bm{v}}^{(r)})$ if $\mu>\lambda_{1}$, $\lambda_{p} > \Vert\bm{h}(\theta_0)\Vert_2^2$ and $\lambda_c\geq \frac{M}{8}\lambda_2+\Vert\tilde{\bm{h}}(z)\Vert_2$ hold, namely, $f^{\prime}(\check{\bm{v}}^{(r+1)})\leq f^{\prime}(\check{\bm{v}}^{(r)})$. 
Since the sequence $\{f^{\prime}(\check{\bm{v}}^{(r)})\}$ yielded by the RNM is monotonically non-increasing and lower bounded, the objective function $f^{\prime}(\bm{v})$ in \eqref{Dinkelbach_v2} converges to a finite value. 
Therefore, the RNM is convergent. 

Based on the convergence of RNM, we have $f^{\prime}(\bm{v}^{(q+1)})\leq f^{\prime}(\bm{v}^{(q)})$.
Then we prove that the sequence of the Dinkelbach variable $\{z^{(q)}\}$ is monotonically non-decreasing and converges to a finite value.

Express the objective function $E(\bm{v})$ in \eqref{Optprob_v} as 
\begin{equation}
	E(\bm{v}) =\frac{A(\bm{v})}{B(\bm{v})}
\end{equation}
where $B(\bm{v})=\sum_{i=1}^{I}\sigma_i^2|\bm{w}^H \bm{\Phi}(\theta_i)\bm{u}|^2+\sigma_n^2\bm{w}^H\bm{w} >0$ and $A(\bm{v})=|\bm{w}^H \bm{\Phi}(\theta_0)\bm{u}|^2 \geq 0$.
At the $q$-th iteration of Algorithm 1, we update the Dinkelbach variable $z^{(q)}$ by using \eqref{Dinkpar_v}. 
Note that $z^{(q)}=A(\bm{v}^{(q)})/B(\bm{v}^{(q)})$. 
We rewrite the problem \eqref{Dinkelbach_v1} as
\begin{equation}
	\label{eq:pdinkelbach2}
	\min_{\bm{v}} \ z^{(q)}B(\bm{v})-A(\bm{v}) \ \ \text{s.t.} \ |v_{m}|=1, \ \forall m.\\
\end{equation}
Recalling that $f^{\prime}(\bm{v}^{(q+1)})\leq f^{\prime}(\bm{v}^{(q)})$, we have
\begin{equation}
	\begin{split}
		z^{(q)}B(\bm{v}^{(q+1)})-A(\bm{v}^{(q+1)}) \leq z^{(q)}B(\bm{v}^{(q)})-A(\bm{v}^{(q)})=0.
	\end{split}
\end{equation}
That is to say, 
\begin{equation}
	\begin{split}
		z^{(q)}=\frac{A(\bm{v}^{(q)})}{B(\bm{v}^{(q)})} \leq \frac{A(\bm{v}^{(q+1)})}{B(\bm{v}^{(q+1)})} = z^{(q+1)}.\\
	\end{split}
\end{equation}

Note that the sequence $\{z^{(q)}\}$ is upper bounded because $E(\bm{v})$ is upper bounded.
Therefore, $\{z^{(q)}\}$ converges to a finite value.
Since $E(\bm{v}^{(q)})=z^{(q)}$, the sequence of $\{E(\bm{v}^{(q)})\}$ also converges to a finite value, namely, Algorithm 1 is convergent.  

\bibliographystyle{IEEEtran}
\bibliography{RIS_TRBF1}
	
\end{document}